\newif\iffulledition
\fulleditiontrue
\documentclass[journal]{IEEEtran}

\usepackage{color}

\usepackage{cite}
\usepackage{url}

\ifCLASSINFOpdf
  \usepackage[pdftex]{graphicx}
\else
  \usepackage[dvips]{graphicx}
\fi

\ifCLASSOPTIONcompsoc
  \usepackage[caption=false,font=normalsize,labelfont=sf,textfont=sf]{subfig}
\else
  \usepackage[caption=false,font=footnotesize]{subfig}
\fi

\usepackage{amsmath}

\usepackage{stfloats}
\usepackage{multirow}

\newlength\figwidth
\ifCLASSOPTIONonecolumn
\else
\fi
\newlength\figheight
\usepackage{pgf}
\usepackage{pgfplots}
\pgfplotsset{
label style={font=\footnotesize},% This does not work, so have to add to the end of each axis' stye.
legend style={font=\footnotesize},
tick label style={font=\footnotesize},
width=\figwidth,
xlabel near ticks,
ylabel near ticks,
scaled ticks=false, % Do not extract common exponents
xticklabel style={/pgf/number format/fixed}, % Do not use scientific format for y-axis
yticklabel style={
/pgf/number format/fixed,
/pgf/number format/precision=5}, % Do not use scientific format for y-axis
}
\iffulledition
\pgfplotsset{
height=\figheight,
}
\else
\pgfplotsset{
height=0.7\figheight,
}
\fi
\begin{document}

\title{``Flow Size Difference'' Can Make a Difference: Detecting Malicious TCP Network Flows Based on Benford's Law}

\author{%
Aamo~Iorliam,
Santosh~Tirunagari,~\IEEEmembership{Student~Member,~IEEE,}
Anthony~T.S.~Ho,~\IEEEmembership{Senior~Member,~IEEE,},
Shujun~Li,~\IEEEmembership{Senior~Member,~IEEE,}
Adrian~Waller,
and~Norman~Poh,~\IEEEmembership{Member,~IEEE}% <-this % stops a space
\thanks{Aamo Iorliam, Santosh Tirunagari, Anthony T.S.\ Ho, Shujun Li and Norman Poh are with the Department of Computer Science, University of Surrey, Guildford, UK.}%
\thanks{Adrian Waller is with Thales UK, Research and Technology (TRT), Reading, UK.}%
\thanks{Corresponding co-authors are Anthony T.S.\ Ho (\protect\url{a.ho@surrey.ac.uk}), Shujun Li (\protect\url{http://www.hooklee.com/}) and Adrian Waller (\protect\url{Adrian.Waller@uk.thalesgroup.com}).}%
\iffulledition\else\thanks{A full edition of this paper is available at \protect\url{https://arxiv.org/abs/1609.04214}.}\fi}

\iffulledition\else
\markboth{IEEE Transactions on Information Forensics and Security,~Vol.~XX, No.~X, 201X}%
{Iorliam \MakeLowercase{\textit{et al.}}: ``Flow Size Difference'' Can Make a Difference for IDS}
\fi

% The following command is needed in second column of first page if using \IEEEpubid.
%\IEEEpubidadjcol


\maketitle

\begin{abstract}
Statistical characteristics of network traffic have attracted a significant amount of research for automated network intrusion detection, some of which looked at applications of natural statistical laws such as Zipf's law, Benford's law and the Pareto distribution. In this paper, we present the application of Benford's law to a new network flow metric ``flow size difference'', which have not been studied before by other researchers, to build an unsupervised flow-based intrusion detection system (IDS). The method was inspired by our observation on a large number of TCP flow datasets where normal flows tend to follow Benford's law closely but malicious flows tend to deviate significantly from it. The proposed IDS is unsupervised, so it can be easily deployed without any training. It has two simple operational parameters with a clear semantic meaning, allowing the IDS operator to set and adapt their values intuitively to adjust the overall performance of the IDS. We tested the proposed IDS on two (one closed and one public) datasets, and proved its efficiency in terms of AUC (area under the ROC curve). Our work showed the ``flow size difference'' has a great potential to improve the performance of any flow-based network IDSs.
\end{abstract}

\iffulledition
\fi

\ifCLASSOPTIONpeerreview
\begin{center} \bfseries EDICS Category: NET-ATTP-INT \end{center}
\fi
\IEEEpeerreviewmaketitle

\section{Introduction}

\IEEEPARstart{T}{he} advances of networking technologies have made the whole world (computes, people and things) far more connected than before, but the increasing connectivity has also lead to more opportunities to malicious attackers who find various ways to launch cyber attacks such as distributed denial of service (DDoS) attacks \cite{DDoS_Survey2013}, botnets \cite{BotnetBook2008}, network worms \cite{Worm_Bookchapter2006}, and phishing attacks (e.g.\ phishing emails and rogue WiFi access points) \cite{PhishingBook2006}. The increasing harm of network attacks have become so severe that such attacks have been becoming more and more frequent and sophisticated \cite{DDoS_BBC, DDoS_500Gbps}.

To reduce and prevent harm that can be caused by network attacks, intrusion detection systems (IDSs) have been studied by researchers and deployed by organizations in real world to detect potential attacks automatically \cite{kruegel2005intrusion, NIDS_Book2009}. IDSs can be classified into network-based IDSs and host-based IDSs, which differ from each other in where the detector is deployed and what activities the detector is monitoring to make alerts. Network-based IDSs are deployed within a network and monitor network traffic passing by to identify potential attacks that may target any hosts of the network, while host-based IDSs are deployed on individual hosts (e.g.\ key servers and individual devices) and monitor local activities (local processes, incoming and outgoing network traffic) to detect signs of suspicious behaviors. This paper focuses on network-based IDSs.

According to how network-based IDSs detect attacks technically, they can be classified into four general categories: misuse-based, anomaly-based, specification-based, and hybrid systems \cite{kruegel2005intrusion, NIDS_Book2009}. Misused-based IDSs are largely based on defined patterns (sometimes called ``signatures'') of known attacks so the detection process is around checking if any known patterns are matched. This approach has the advantage of producing very few false positives but it cannot be generalized to cover unknown attacks. On the other hand, anomaly-based IDSs are based on learned normal activity profiles of the network traffic (for network-based IDSs) and the system (for host-based IDSs), so anything deviating from normal behavior will be labelled as suspicious, which can cover totally unknown attacks but also unusual non-malicious activities. Specification-based IDSs are very similar to anomaly-based IDSs but here the normal profiles are specified rather than learned. As its name implies, hybrid IDSs combine different approaches (typically a misused-based component for known attacks and an anomaly-based component for unknown attacks) to overcome limitations of other simple approaches. Some researchers prefer other terms e.g.\ Debar et al.\ suggested using knowledge- and behavior-based IDSs for misuse- and anomaly-based IDSs \cite{Debar_IDS_taxonomy2000}, and Dra{\v{s}}ar et al.\ proposed to classify (flow-based) IDSs based on the concept of similarity \cite{dravsar2014similarity}.

According to the network traffic data used, network-based IDSs can also be classified into packet- and flow-based approaches \cite{dreger2004operational, bejtlich2004tao, sperotto2011flow}. Packet-based IDSs look at all packets passing by in the network traffic, but flow-based IDSs analyze characteristics of network flows (a sequence of network packets). Flow-based IDSs are computationally far more efficient, but have a coarser temporal resolution in locating attacks and can miss important information related to application-level attacks. Flow-based IDSs also have the advantage that they can adapt to different/evolving network infrastructure more easily and are considered more privacy-friendly as packet content recording and inspection are avoided \cite{steinberger2013anomaly}.

A lot of network-based IDSs have employed statistical approaches to detecting signs of attacks, and some natural statistical laws such as Zipf's law, Benford's law, the Pareto distribution and the Weibull distribution \cite{newman2005power, Tao2009ThreeLaws, BenfordsLawBook2015, WeibullDistribution} have been explored to generate statistical features for IDS purposes \cite{yegneswaran2003internet, Hamadeh_MScThesis2004, Wang2004, Cai2007, Gu_USENIX_Security2007, Kim_AINAW2007, YeZheng2011, Arshadi_WeibullIDS_EMS2011, yu2012discriminating, arshadi2014empirical}. The term ``natural'' refers to the fact that many natural processes often follow them while artificially created ones tend to not. Since attacks are normally artificially crafted and mostly generate ``unnatural'' network traffic, those natural laws can often form the basis (or part) of an anomaly-based IDS.

In this paper, we report our observation that a new network flow metric we call ``flow size difference'' follows Benford's law closely for normal TCP flows but not for malicious ones. We then propose a new method of applying Benford's law to this new metric to build an unsupervised network-based IDS to detect malicious TCP flows. In addition to having all merits of flow-based IDSs, our proposed IDS does not require any training and it provides two simple operational parameters with a clear semantic meaning so that the IDS operator can set and tailor the parameters intuitively to adjust the performance of the proposed IDS. The main limitation of the proposed method is that it requires a relatively large time window of flows (to provide statistical confidence) so its temporal resolution is relatively low, but this can be overcome by combining it with other IDSs with a higher temporal resolution (e.g.\ using the proposed method to detect regions of interest which are then further analyzed using other methods). The proposed IDS was inspired by our observations on a large number of TCP flow datasets, and its performance was verified with two (one closed and one public) datasets.

The rest of this paper is organized as follows. In the next section, we overview related work on Benford's law and other related natural laws and statistical IDSs. Section~\ref{sec:flow_def} gives a brief explanation to the concept ``network flow'' and explains how we define TCP flows in our work. In Sec.~\ref{sec:OurMethod} we introduce the basic concept behind the new metric ``flow size difference'' and explain how Benford's law can be applied to measure (un)naturalness of TCP flows to distinguish malicious flows from normal ones. Section~\ref{sec:Experiments} presents some experimental evidence on the usefulness of Benford's law based naturalness detector with a number of selected network traffic datasets with normal, malicious and mixed TCP flows. Then, in Sec.~\ref{sec:IDS} we report the experimental results of using the proposed naturalness detector to build a simple and fast threshold-based IDS, which was applied to one closed and one public datasets to show its potential as a new element for designing network IDSs. We discuss operational advantages and limitations of the proposed methods and how it can be combined with other methods in Sec.~\ref{sec:discussion}, with our planned future work. Finally, the last section concludes the paper.

\section{Related Work}
\label{sec:RelatedWork}

\subsection{Benford's Law}

The first known description of Benford's law was given by Simon Newcomb in 1881 \cite{Newcomb1881}, but Newcomb's work was largely forgotten and later ``rediscovered'' and studied more systematically by Frank Benford in 1938 \cite{Benford1938}.\footnote{Some researchers prefer calling the law Newcomb-Benford Law (NBL) \cite{Formann2010BenfordLaw}, but Benford's law remains the mostly used term. As a matter of fact, in \cite{Newcomb1881} Newcomb actually implied that the law as an approximate phenomenon had been known to most people who frequently used logarithmic tables.} The simple form of the law states that the first digits of many real-life sets of numerical data follow the following probability distribution:
\begin{equation}
\label{eqn:ben}
P_d = \log_{10}\left(1+d^{-1}\right),
\end{equation}
where $d=1,\ldots,9$ (each possible value of the first digit) and $P_d$ denotes the probability of $d$. This simple form can be generalized to other bases other than 10 and other significant digits other than the first one \cite{BergerHill2011BenfordSurvey, BenfordsLawBook2015}.

Researchers have proposed different explanations to why Benford'a law arises in real-life datasets \cite{Pinkham1961BenfordLaw, Hill1995BenfordLaw, Hill1995}. One general agreement is that Benford's law applies to numbers with a scale-invariant distribution \cite{Pinkham1961BenfordLaw, Hill1995BenfordLaw}, but a complete and convincing theoretical explanation remains unsettled due to some mathematical subtleties as described in \cite{BergerHill2011BenfordLaw}.

Although Benford's law holds for many real-life datasets, it is not a universal law as there are many exceptions. There are also some other statistical laws / distributions which are closely linked to Benford's law, such as the Stigler distribution \cite{Lee2010StiglerLaw}, Zipf's law, the Pareto distribution \cite{irmay1997relationship, newman2005power, Tao2009ThreeLaws}, the Weibull distribution \cite{WeibullDistribution} and a generalized Benford's law with two added tunable parameters \cite{Fu}.

Due to the fact that real-life datasets in many different fields follow Benford's law, it has found diverse applications in many applications reflected from the huge number of publications spreading across many disciplines \cite{BeebeBenfordBibTeX, benfordonline}. Particularly, Benford's Law has been one of the most-studied natural laws for fraud detection \cite{Durtschi2004BenfordLaw4FraudDetection, DiekmannJann2010BenfordLawFraudDetection, Nigrini2012BenfordLawBook} and multimedia forensics \cite{Acebo, Fu, li2008detecting, li2012detection, Qadir}, which have close links to IDSs. Some researchers also proposed some generalized forms of Benford's law to adapt to solve practical problems, e.g., Fu et al.~\cite{Fu} proposed a generalized Benford's law with two tunable parameters to fit the multimedia forensics data they were studying.

\subsection{IDS}

We have briefly discussed the general classification of IDSs in the Introduction, and here we will focus mainly on statistical methods particularly those based on natural statistical laws and applied to build anomaly-based IDSs.

Theoretically speaking, statistics play a foundational role for all IDSs (even for methods not based on statistics directly) because an IDS's performance is essentially based on statistics of normal and malicious activities \cite{Helman_IDS_Foundations_CSFW92}. As a matter of fact, there is an implicit or explicit assumption that attack activities are statistically different from normal ones \cite{NIDS_Book2009}. Another example is that all IDSs face a well-known issue called ``base-rate fallacy'': the probability of any malicious event takes place in the whole range of activities is generally very small and the benign activities almost always dominate over a relatively long period of time, which can lead to foundational difficulties in the design, development and testing of IDSs \cite{Axelsson_BaseRateFallacy_CCS99}, e.g.\ it can be practically difficult to get sufficient attack data to train and test an IDS based on supervised learning methods \cite{SommerPaxson_IDS2010, dravsar2014similarity}.

To apply any statistical methods, one needs to have at least one random variable which can be observed over time and modeled by the IDS. In his seminal paper on anomaly-based IDSs \cite{DenningIDS1987} Denning called such random variables \emph{metrics}, which will be the term we follow in this paper. The metric(s) chosen should allow normal and malicious activities to differ from each other statistically so that two different statistical models can be built to support classification. The observed \emph{samples} of a selected metric may be used directly for classification purposes, but they can also be aggregated to derive other higher-level indicators (e.g.\ mean, standard deviation and entropy) before the classification step takes place. Here we use the term \emph{features} to denote those indicators fed into a classification system for making decisions, which is a standard term used by the machine learning community to call data feed into a (statistical or non-statistical) classifier.

While the statistical models of selected metrics do not have to be based on any assumed distributions (as argued by Denning \cite{DenningIDS1987}), many researchers noticed some metrics taken from non-malicious network traffic follow some known distributions closely but those from malicious deviate significantly \cite{yegneswaran2003internet, Hamadeh_MScThesis2004, Wang2004, Cai2007, Gu_USENIX_Security2007, Kim_AINAW2007, YeZheng2011, Arshadi_WeibullIDS_EMS2011, yu2012discriminating, arshadi2014empirical}. Those distributions studied include Zipf's law, the Pareto distribution, the Weibull distribution and also Benford's law.

Surprisingly, despite the fact that Benford's Law has been widely used for fraud detection and digital forensics, there has been very limited work on applying Benford's law to IDS. The first work we are aware of was done by Hamadeh in his 2004 master's thesis \cite{Hamadeh_MScThesis2004}. In his work he looked at three candidate metrics: the number of hits of webserver, the number of bytes transmitted in the response, and the inter-arrival time between two consecutive visits. He observed that the inter-arrival time followed Benford's law better than the other two. Hamadeh mentioned the possibility of applying the observation to an IDS, but fell short of actually conducting any experiment. Note that Hamadeh's work is not network flow based, but around activities of visiting a webserver. The only another work (and the only work on flow-based IDS) we could find is due to Arshadi and Jahangir \cite{arshadi2014benford} who showed that the inter-arrival time of two consecutive normal TCP flows Benford's law closely so an IDS can be built on top of this fact. They conducted some manual inspection on selected malicious flows, but did not build an actual IDS so its actual performance is unclear. Arshadi and Jahangir also studied the source of Benford's law and attributed it to the fact that normal TCP flows' inter-arrival time closely follows the Weibull distribution, which can derive Benford's law. In \cite{arshadi2014empirical}, Arshadi and Jahangir also studied using the Weibull distribution with the inter-arrival time for IDS purposes, and provided some results on the actual performance of such an IDS.

\section{Network Flows and TCP Flows}
\label{sec:flow_def}

Before discussing our work in greater details, we clarify a bit more about what we mean by ``TCP flows'' and ``network flows'' in general. IDS researchers do not have a consistent understanding on these two terms and different flow-based IDSs can use very different definitions of network flows.

Loosely speaking, a (network) flow, sometimes called a session, a stream, or a conversation, is a sequence of network packets sharing some common criteria such as two end-point IP addresses over the Internet \cite{Fredj2001flow, bejtlich2004tao, tune2013internet, Hofstede2014flow}. In principle, the concept can be applied to any network protocol at any layer of a network. This concept has been widely used for network traffic monitoring purposes, and a specific protocol called NetFlow working at the IP layer was popularized by Cisco due to its widely deployed routers, which was later developed into IPFIX (Internet Protocol Flow Information Export) \cite{Hofstede2014flow}. A flow does not necessarily correspond to a single communication session as defined by the underlying protocol, e.g.\ Fredj et al.\ \cite{Fredj2001flow} defined a flow as a collection of packets transmitting a complete document (such as a web page) and one definition Barakat et al.\ used in \cite{Fredj2001flow} is ``a stream of packets having the same/24 destination address prefix''.

While there are many different ways to define a network flow, one of the most common definitions uses the following five criteria: source and destination IP addresses, source and destination port numbers (0 for protocols that do not use ports), and protocol type. When we talk about ``source'' and ``destination'' it is clear that the flow defined is unidirectional, which is the case for NetFlow and IPFIX flows. However, as Bejtlich pointed out in \cite{bejtlich2004tao}, connection-oriented protocols (e.g.\ TCP) are more suited to be represented as a flow as compared to connectionless protocols (e.g.\ IP, UDP and ICMP). This is because the former are structured in such a way that there exists a clear beginning, middle, and end to a flow, whereas the latter are not structured around the concept of connections so often time expiration conditions have to be used to \emph{arbitrarily} set flow boundaries. For connection-oriented protocols, each connection corresponds to a bidirectional flow which is the merge of two unidirectional flows.

In flow-based IDSs different flow definitions are used. IP flows defined following NetFlow or IPFIX specifications are widely used \cite{sperotto2010overview} since such flow information can often be obtained directly from routers and supported by open-source tools such as NFDUMP and NfSen\iffulledition\ \cite{NFDUMP, NfSen}\fi. Some researchers focused on bidirectional TCP flows only \cite{arshadi2014benford}, which can be justified by the fact that most IP traffic over the Internet is TCP traffic (e.g.\ in \cite{shiravi2012toward} the ratio was reported to be 95.112\%).

In this paper, we also focused on bidirectional TCP flows as Arshadi and Jahangir did in their work on \cite{arshadi2014benford} considering the dominance of TCP flows on the Internet and the fact that network flows are less well-defined for IP and other connectionless protocols. Note that selecting TCP flows naturally cover all application-layer protocols based on TCP (such as HTTP, the dominating protocol at the application layer \cite{shiravi2012toward}).

When working with a pcap file captured by \textsl{libpacap} or \textsl{WinPcap}, the TCP flows we work with can be generated using the following command line with \textsl{tshark} (part of \textsl{WireShark}):\iffulledition

\texttt{tshark -r <input pcap file> -q -z conv,tcp > <output file>}.

\else\ \texttt{tshark -r <input pcap file> -q -z conv,tcp > <output file>}.\fi\ The output file will then contain a list of ``TCP conversations'' which is a term used by \textsl{WireShark} to denote TCP flows.\iffulledition\footnote{\textsl{WireShark} uses another term ``TCP stream'' to denote the payload of a ``TCP conversation''. To avoid confusion we will use the term ``TCP flow'' consistently in this paper.} In \textsl{WireShark}'s user interface, TCP flows of a given network traffic dataset can be obtained via the menu item ``Statistics'' $\rightarrow$ ``Conversation'' and then click ``TCP'' tab. Then the TCP flows shown in the interface can be exported using ``Copy'' button at the bottom.\fi\ The TCP flows obtained this way contain the following attributes (those in boldface are essential for our work): source IP address, source port number, destination IP address, destination port number, \textbf{total number of packets transferred between source and destination}, \textbf{total number of bytes transferred between source and destination}, packets transferred from source to destination, bytes transferred from source to destination, packets transferred from destination to source, bytes transferred from destination to source, \textbf{relative start time} (as a timestamp, 0 = the beginning of the whole network traffic), \textbf{duration}, source-to-destination bitrate (bit per second), and destination-to-source bitrate. When working with other network traffic data, as long as we can extract the above attributes of each TCP flow in boldface, our proposed method will work without any problem.

\section{Our Proposed Method}
\label{sec:OurMethod}

In this section we will describe the basic concepts behind the IDS based on Benford's law and the new metric ``flow size difference'' and different aspects we need to consider around the proposed IDS.

\subsection{IDS Structure and Component}

We consider the typical structure of an anomaly-based IDS working with TCP flows. It starts with one or more selected metrics which should follow Benford's law closely for normal TCP flows but deviate from it significantly and consistently for malicious ones. Then, a number of samples of each selected metric will be collected (one per TCP flow) to detect any significant deviation from Benford's law, where we will use a sliding window to cover a sufficient number of TCP flows (i.e., samples of the metric) so that any deviation can be detected with high confidence and an acceptable temporal resolution (which will help to tailor the detection accuracy as well especially to balance false positive and false negative rates). The flow window size $W$ will be an important system parameter, and it can be fixed or dynamically adapted depending on the nature of the networking environment being monitored. The deviation will be calculated based on a similarity metric between the Benford's law and the actually observed distribution, which is used as a feature for a binary classifier to classify each flow window into two classes: normal (non-malicious), or attack (malicious).

If we have more than one usable metric, each one can be used alone or they can be combined to inform a classifier handling multiple features or the alerts generated from each metric are pooled to derive a single decision. In this paper, we look at a simpler setting: each usable metric is used alone and a simple threshold $T$ is used to construct a binary classifier. We go for the simpler setting because our main goal here is to identify \textbf{new} metrics that can work with any IDS to improve its performance rather than to produce yet another IDS competing with other systems. We actually do not anticipate our IDS will outperform many more complicated IDSs due to its simplistic structure. To some extent, the proposed IDS structure is used to set up a proper context so that we can test the usefulness of the proposed new metrics in more complicated IDSs. In our future work we will look at more complicated settings and how the identified new metrics can be used to improve other IDSs. We will discuss more about these issues in Sec.~\ref{sec:discussion}.

\subsection{The Metrics}

As mentioned above, the first and the most important component of our IDS is at least one metric that can work with Benford's law. Since we aim at identifying new metrics, we do not look at the inter-arrival time already studied by other researchers \cite{arshadi2014benford}. Looking at the attributes of a TCP flow listed at the end of Sec.~\ref{sec:flow_def}, we identified two candidate metrics that have not been previously studied for IDS:

\subsubsection{Flow size}: The flow size distribution has been studied extensively in the networking literature and estimation of flow size distribution has been an active research topic \cite{Kumar2004FSDestimate, Tune2014OFSS}. It has been known that the flow size distribution is typically long-tailed and its exact form heavily depends on the underlying protocol and the networking environment \cite{Garsva2015FlowSizeDistribution}. Flow size distribution has been been studied in IDS research but mostly for entropy-based approaches where the entropy of the observed distribution is calculated as a feature for detection \cite{Nychis2008entropyIDS, Basicevic2016FSD_IDS}. We were unaware of work linking flow size distribution to Benford's law so found it interesting to investigate flow size as a potential metric for applying Benford's law to IDS.

\subsubsection{Flow size difference}: In addition to flow size, we also noticed that the ``flow size difference'', which is defined as the numeric difference of two consecutive TCP flows' sizes, seems to be another potential metric of interest because it inherits some features of flow size (e.g.\ long-tailedness) but differs significantly from the flow size itself. We did not find any work on the distribution of flow size difference of TCP flows or on its application in IDS\footnote{If we know the flow size distribution, it is possible to derive the distribution of the flow size difference assuming the flow size is an i.i.d.\ sequence which is unfortunately not the case for most cases.}, which is not totally surprising since the flow size difference does not seem to be obviously useful for network traffic analysis and management purposes. However, our experiments revealed that it seemed to follow Benford's law well, so we added it as a candidate. For the flow size difference, we will ignore the sign bit so the metric we are considering here is actually the absolute value of the flow size difference. In the following, we will simply use the term ``flow size difference'' to denote the absolute value unless otherwise stated.

Note that the flow size can be defined by bytes or packets, so we actually have two different variants for each of the above two candidate metrics. These two variants are not linked but cannot be directly derived from each other since the packet size (in byte) varies over time and cross applications.

\subsection{From Flows to Flow Windows}
\label{sec:FlowWindow}

Since Benford's law is a distribution, we need to collect enough samples of a given metric to be able to construct an observed distribution which then can be compared against the target distribution for detecting any deviation. This is why we need to have a flow window for any distribution-based IDSs, which is a well-known fact. Since the Benford's law distribution has only 9 values, it can be expected that the minimum flow window size will not be very large so that the temporal resolution of the IDS will not be too low. We will report our experimental results in Sec.~\ref{sec:Experiments} on how the value of $W$ can be determined. In addition to the parameter flow window size $W$, there is also another one ``window sliding step'' $S$ which defines how much the flow window slides at one time. In principle $S$ can range from 1 to $W$, and taking a smaller value can potentially help refine the temporal resolution to some extent. In our work, we take $S=W/2$ as a representative value.

\subsection{Flow Ordering}
\label{sec:FlowOrdering}

For the flow size difference, changing ordering of flows will obviously change each sample of the metric and thus its observed distribution. For the flow size, the flow ordering will not make a direct difference, but can influence what flows are included in each flow window thus influence the observed distribution. It is therefore a valid question to ask if such orderings will make a difference and if so how we should order flows. Looking at all the attributes of a TCP flow, we selected the following four typical ordering options for consideration:
\begin{enumerate}
\item (Start Time, End Time)
\item (End Time, Start Time)
\item (Source IP address, Destination IP address, Start Time)
\item (Source IP address, Source Port number, Destination IP address, Destination Port number, Start Time)
\end{enumerate}
Each attribute in the above ordering option is considered from the left to right. For flows with the same values for all ordering criteria, the original order in the raw network traffic log will be kept. Since Benford's law is a natural law and scale-invariant, we hypothesized that the ordering should not have a big impact on the compliance with (for normal flows) or deviation from (for malicious flows) Benford's law. Our experimental results will be given in Sec.~\ref{sec:Experiments}.

\subsection{Zero Handling}

For the flow size difference we have to consider a special issue: zero is now a possible value since two consecutive flows can have the same size (which is highly likely in some attacks such as DDoS and port scanning). Since Benford's law does not actually cover the digit zero, we need to handle such zero flow size differences properly to avoid system crashing and false detections. There are several options we can consider: 1) skipping all zeros; 2) extending Benford's law to cover digit zero, i.e., simply add $P(0)=0$ into the target distribution so now we have 10 digits to consider (but 0 should never appear). The first option is trivial to implement but may lose important information about a specific pattern (repeated flows of the same size) around some attacks. The second option looks trivial to implement but has one non-trivial issue regarding how we can measure goodness of fit and deviation from Benford's law which will be discussed in the next subsection.

\subsection{Measuring Goodness of Fit / Deviation}
\label{sec:GoodnessMetrics}

To check the goodness of fit to and deviation from the target distribution described by Benford's law, we need a proper similarity metric\iffulledition\footnote{In this paper we use the term ``metric'' loosely without following the rigorous definition of metrics in mathematics.}\fi. One of the most commonly-used metrics in applications of Benford's law is the $\chi^2$ divergence \cite{Acebo, Fu, li2008detecting, li2012detection, Qadir} which can be defined for Benford's law case as follows:
\begin{equation}\label{eq:chi2divergence}
\chi^2 = \sum_{d=1}^9 \frac{(\hat{P}_d-P_d)^2}{P_d},
\end{equation}
where $P_d$ is the probability of first digit $d$ defined in Benford's law as in Eq.~\eqref{eqn:ben} and $\hat{P}_d$ is the actually observed probability of $d$. It is effectively the test statistic of Pearson's $\chi^2$ test \cite{Plackett1983chi2test} against the target distribution.

In the IDS context, there are many other similarity metrics one can use for the goodness of fit / deviation test. In a 2015 survey \cite{weller2015survey} Weller-Fahy et al.\ gave a very comprehensive overview of this topic, classifying different metrics used in network-based IDSs into four categories: 1) power distances (e.g.\ Euclidean distance and Manhattan distance), 2) distances on distribution laws (e.g.\ $\chi^2$ divergence, Kullback-Leiber divergence, and entropy-based metrics), 3) correlation similarities (e.g.\ Spearman $\rho$ rank correlation, Kendal $\tau$ rank correlation, Pearson product-moment correlation coefficient, and the cosine similarity), and 4) other metrics (e.g.\ dice similarity and Geodesic distance).

In our initial experiments investigating the potential of Benford's law (as reported in Sec.~\ref{sec:Experiments}), we stuck to the $\chi^2$ divergence. In the performance evaluation of the IDS we also tested the following metrics in order to find out if there are significant differences when different metrics are used: Euclidean distance, Manhattan distance, Pearson product-moment correlation coefficient, and cosine similarity. Although all similarity metrics are well-known, we list them below for the sake of completeness (all tailored for Benford's law following notations in Eq.~\eqref{eq:chi2divergence}).
\begin{itemize}
\iffalse
\item
Squared $\chi^2$ divergence:
\[
\hat{\chi^2} = \sum\nolimits_{d=1}^9 \frac{(\hat{P}_d-P_d)^2}{\hat{P}_d+P_d}
\]
\fi

\item
Euclidean distance:
\[
\text{ED} = \sqrt{\sum\nolimits_{d=1}^9\left(\hat{P}_d-P_d\right)^2}
\]

\item
Manhattan distance:
\[
\text{MD} = \sum\nolimits_{d=1}^9\left|\hat{P}_d-P_d\right|
\]

\item
Canberra distance:
\[
\text{CD} = \sum\nolimits_{d=1}^9\frac{\left|\hat{P}_d-P_d\right|}{\hat{P}_d+P_d}
\]

\item
Pearson product-moment correlation linear coefficient ($\overline{x}$ denotes the mean of $x$):
\[
\text{CC} =
\frac{\sum_{d=1}^9\left(\hat{P}_d-\overline{\hat{P}_d}\right)\sum_{d=1}^9\left(P_d-\overline{P_d}\right)}
     {\sqrt{\sum_{d=1}^9\left(\hat{P}_d-\overline{\hat{P}_d}\right)^2}\sqrt{\sum_{d=1}^9\left(P_d-\overline{P_d}\right)^2}}
\]

\item
Cosine similarity:
\[
\text{CS} =
\frac{\sum_{d=1}^9\left(\hat{P}_dP_d\right)}
     {\sqrt{\sum_{d=1}^9\hat{P}_d^2}\sqrt{\sum_{d=1}^9P_d^2}}
\]
\end{itemize}

When Benford's law is extended to cover $d=0$, some of the above metrics can still be used by just changing the starting index $d=1$ to 0. For metrics where $P(d)$ appears as the denominator (i.e., $\chi^2$ divergence), we have to skip $P(0)$ as this can cause singularity, but this does not mean the effect of zero flow size differences are ignored as a non-zero $\hat{P}_0$ will reduce the probability of all other digits so can still lead to a difference in the similarity metric calculated.

In addition to the above metrics, we also considered Kullback-Leiber divergence since it is also widely used to measure differences between distributions. Kullback-Leiber divergence is actually not a proper similarity metric for the purpose of this study because the underlying assumption, $\forall d$, $P_d=0\to\hat{P}_d=0$, is not guaranteed to hold in real world. In addition, although $d=0$ is not counted in Benford's law, it can appear for the flow size difference and in the extreme case one can have $\hat{P}_0=1,\hat{P}_1=\cdots=\hat{P}_9=0$ (which does happen for some datasets). Therefore, we propose to use a modified edition of Kullback-Leiber divergence by introducing a new term to count the effect of first digit 0 more properly:
\begin{equation}
\text{KLD}^* = \hat{P}_0\theta_{\text{KLD}}+\sqrt{\sum\nolimits_{d=1}^9\hat{P}_d\log_2\left(\frac{\hat{P}_d}{P_d}\right)},
\end{equation}
where $\theta_{\text{KLD}}$ is a parameter representing the unit contribution of digit 0 to the final divergence value, i.e., representing the ill-defined term $\log_2(\hat{P}_0/P_0)=\infty$. In our experiments, we used $\theta_{\text{KLD}}=2\log_2(1/P_9)$ following the heuristic assumption that the contribution of $\hat{P}_0=1$ to the divergence is twice as large as that of $\hat{P}_9=1$. This proved sufficient to make the modified KLD a good metric for all cases of our IDS.

\section{Initial Experiments}
\label{sec:Experiments}

In this section, we describe some initial experiments we conducted to test the two types of new metrics (flow size and flow size difference), leading to the conclusion that the flow size difference is a promising new metric for IDS. The performance evaluation of the IDS based on the flow size difference will be reported in Sec.~\ref{sec:IDS}.

\subsection{Network Traffic Datasets Used}

To properly test the new metrics, we used three groups of network traffic datasets in our experiments. The first group are datasets that do not contain known malicious traffic (so considered normal/non-malicious). The second group are datasets containing known malicious traffic. The third group are datasets with mixed traffic with both malicious and non-malicious flows. All datasets were pre-processed using \textsl{tshark} to get bidirectional TCP flows.

Non-malicious datasets we considered include:
\begin{enumerate}
\item
LBNL/ICSI dataset \cite{LBNLdata}: We used the whole dataset (442,623 TCP flows).
% This has packets spanned for more than 100 hours of activity from a total of several thousand internal hosts.

\item
UNB ISCX 2012 intrusion detection evaluation dataset \cite{ISCX2012dataset, shiravi2012toward}: We used two files with non-malicious traffic on Friday 11 June, 2010 (297,398 TCP flows) and on Wednesday 16 June, 2010 (434,674 TCP flows).

\item
TLERH dataset described in \cite{szabo2010traffic}: We used a file of this dataset obtained from the authors of \cite{szabo2010traffic} (with 16,321 TCP flows).
% This has captured data volume of 6GB, for a duration of 43 hours that contains 12 million packets.

\item
Labeled network flow data from the University of Twente \cite{TwenteIPOM2009Dataset, sperottoIPOM2009}: We used non-malicious data traffic in this dataset (12 separate files, in total 1,373,786 TCP flows).
\end{enumerate}

Malicious datasets we considered include:
\begin{enumerate}
\item
Capture the hacker 2013 competition dataset \cite{CtH2013NAPENTHES, CtH2013HONEYBOT, CtH2013AMAZON}: This dataset contains three separate PCAP files called NAPENTHES, HONEYBOT, and AMAZON, which contain 5,440 TCP flows in total.

\item
2009 Inter-Service Academy CDX dataset \cite{CDX_2009}: We used 502,412 TCP flows from this dataset.
% This includes packet captures generated by National Security Agency (NSA) Red Team activity.

\item
MACCDC dataset \cite{MACCDC_netresec}: We randomly picked 4 files from this dataset (with 11807, 65536, 9765 and 23949 TCP flows, respectively).
% This data has normal flows with periodic attacks from a volunteer Red Team.

\item
Labeled network flow data from the University of Twente \cite{TwenteIPOM2009Dataset, sperottoIPOM2009}: We used all malicious traffic in this dataset (13,276,385 TCP flows).
% collected from a honeypot

\item
Traffic data from Kyoto University's honeypots \cite{KyotoHoneypotsDataset, song2008cooperation}: We used four files from this dataset (recorded on 1 to 4 November, 2006, with 11807, 124816, 9765, and 23494 flows, respectively).
% Are these flows are TCP flows? Not totally clear from the description.
\end{enumerate}

We used two datasets with mixed traffic:
\begin{itemize}
\item
UNB ISCX 2012 intrusion detection dataset \cite{ISCX2012dataset, shiravi2012toward}: We used mixed traffic on five dates: 1) 12 June, 2010 (95,177 TCP flows including 2,082 malicious flows); 2) 13 June, 2010 (221,026 TCP flows including 20,358 malicious flows); 3) 14 June, 2010 (122,298 TCP flows including 3,771 malicious flows); 4) 15 June, 2010 (441,563 TCP flows including 37,378 malicious flows); and 5) 17 June, 2010 (329,378 TCP flows including 5,203 malicious flows).

\item
ISOT dataset \cite{ISOTdataset, saad2011detecting}: We used 40,000 flows extracted from the dataset.
\end{itemize}

\subsection{Flow Size or Flow Size Difference?}

The first initial experiment we conducted is to determine if both flow size and flow size difference are promising metrics for IDS purposes. We use one dataset from each of the three groups to gain insights about their potentials. Table~\ref{tab:flowsizevsflowsizedifference} summarizes the results, which show that the flow size difference seems to follow Benford's law better than the flow size although for both cases the expected order of the $\chi^2$ divergence values were observed: non-malicious $<$ mixed $<$ malicious. We also repeated this experiment with some other datasets and observed largely the same pattern, therefore we decided to chose the flow size difference for other experiments. The potential of using the flow size is not totally excluded but we leave it for our future work.

\begin{table*}[!htb]
\caption{$\chi^2$ divergence values for the flow size vs.\ the flow size difference as the metric.}
\label{tab:flowsizevsflowsizedifference}
\centering\small
\begin{tabular}{c|c|c}
\hline
\multirow{2}{*}{Dataset (File)} & \multicolumn{2}{c}{Metric}\\
\cline{2-3}
& Flow Size & Flow Size Difference\\
\hline
Non-malicious traffic: LBNL/ICSI & 0.02577 & 0.00088\\
\hline
Malicious traffic: Capture Hacker (HONEYBOT) & 0.12128 & 0.03999\\
\hline
Mixed traffic: ISCX 2012 (13 June, 2010) & 0.04633 & 0.00603\\	   			   			   		
\hline
\end{tabular}
\end{table*}

\subsection{How to Determine Flow Window Size?}

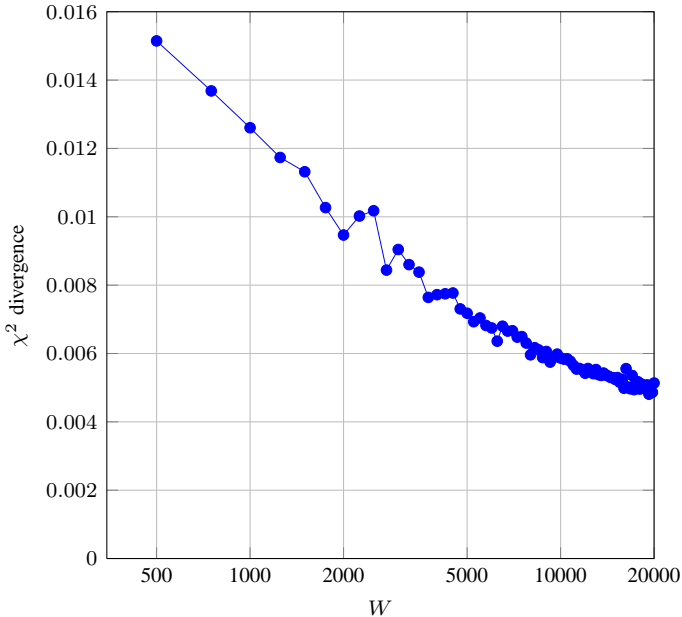
\begin{figure}[!htb]
\centering
% Minimal pgfplots version: 1.3
%  http://www.mathworks.com/matlabcentral/fileexchange/22022-matlab2tikz
\begin{tikzpicture}

\begin{semilogxaxis}[%
xmax=20000,
xlabel={$W$},
ymin=0,
ymax=0.016,
grid=major,
xtick={500, 1000, 2000, 5000, 10000, 20000},
xticklabels={500, 1000, 2000, 5000, 10000, 20000},
ylabel={$\chi^2$ divergence},
label style={font=\footnotesize}% This has to appear after xlabel and ylabel option to work. Seems to be a bug.
]
\addplot[color=blue,solid,line width=0,mark=*,mark options={solid}]
  table[row sep=crcr]{%
500 0.0151463\\
750 0.0136829\\
1000 0.0126058\\
1250 0.0117359\\
1500 0.0113175\\
1750 0.0102689\\
2000 0.00946631\\
2250 0.0100226\\
2500 0.0101766\\
2750 0.00844035\\
3000 0.0090429\\
3250 0.00860065\\
3500 0.00838099\\
3750 0.00764189\\
4000 0.00772135\\
4250 0.00774679\\
4500 0.00777011\\
4750 0.00730508\\
5000 0.00717858\\
5250 0.00692996\\
5500 0.00703949\\
5750 0.0068203\\
6000 0.00674707\\
6250 0.00636127\\
6500 0.00679897\\
6750 0.00665615\\
7000 0.00666609\\
7250 0.00647902\\
7500 0.00649512\\
7750 0.0063058\\
8000 0.00596137\\
8250 0.00617021\\
8500 0.0061057\\
8750 0.00588085\\
9000 0.00605979\\
9250 0.00574775\\
9500 0.00588262\\
9750 0.00598222\\
10000 0.00586625\\
10250 0.00583563\\
10500 0.00584426\\
10750 0.00577105\\
11000 0.00565255\\
11250 0.00554454\\
11500 0.00556232\\
11750 0.005522\\
12000 0.00542198\\
12250 0.00556079\\
12500 0.00547789\\
12750 0.00540942\\
13000 0.00553118\\
13250 0.00538165\\
13500 0.00535907\\
13750 0.00542734\\
14000 0.00536573\\
14250 0.00533853\\
14500 0.00529821\\
14750 0.0052924\\
15000 0.00524186\\
15250 0.00529605\\
15500 0.00515929\\
15750 0.00524382\\
16000 0.00498363\\
16250 0.00555731\\
16500 0.00501738\\
16750 0.00496803\\
17000 0.00535703\\
17250 0.00494139\\
17500 0.00516243\\
17750 0.00517355\\
18000 0.00495959\\
18250 0.00510241\\
18500 0.00501721\\
18750 0.00497707\\
19000 0.00508072\\
19250 0.00480963\\
19500 0.00504801\\
19750 0.00486437\\
20000 0.00513616\\
};
\end{semilogxaxis}
\end{tikzpicture}% 
\caption{Change of $\chi^2$ divergence w.r.t.\ the flow window size $W$.}\label{fig:app_win_size}
\end{figure}

As we discussed in the previous section, the flow window size $W$ is a parameter of the IDS which can allow some level of performance control. While we expected that $W$ does not need to be too large, we need some indication of how we can set it. For a selected dataset (LBNL/ICSI dataset), we calculated the average $\chi^2$ divergence value of all flow windows when $W$ changes from 500 to 20,000, increasing with a step size of 250. Figure~\ref{fig:app_win_size} shows the results. As can be seen, the average $\chi^2$ divergence value decreases rapidly initially but the decreasing rate keeps dropping while $W$ increases. Note that the initial divergence value is also very small (below 0.02) so the results suggest that $W$ can probably be set to a value between several hundreds or thousands in most cases.

\subsection{Flow Ordering}

Another experiment was conducted to check if the flow ordering can change how the flow size difference follows Benford's law. For non-malicious flows in the LBNL/ICSI dataset, the average $\chi^2$ divergence values for all four ordering options listed in Sec.~\ref{sec:FlowOrdering} are: 0.0094497, 0.0086003, 0.0087495 and 0.0073707, all below 0.01. For malicious flows in the HONEYBOT dataset, the average $\chi^2$ divergence values are 0.04102922, 0.039240332, 0.039630320 and 0.036690744, all above 0.03. Repeated experiments on other malicious and non-malicious datasets gave similar results, so we can see that the flow ordering does not really change the $\chi^2$ divergence value much, thus suggesting that the flow size difference is a robust metric against flow ordering.

\subsection{Other Settings}

Our experiments on the two different zero handling options suggested that Benford's law largely works in both cases, although there are indications that for skipping zeros can lead to more fluctuations of the divergence value. Similarly, for the two different variants of the flow size difference metric (flow size as number of byte and number of packet), we observed that both worked reasonably well with Benford's law. Further testing of these settings can be better done via the performance evaluation on the IDS, which we report in Sec.~\ref{sec:IDS}.

\subsection{More Results on Different Datasets}

The above experiments clarified that the flow size difference seems a good metric working with Benford's law. To test its potential further, we ran a large-scale experiment to calculate $\chi^2$ divergence values for a large number of TCP flow windows of many datasets in all the three groups. We used two representative window size (2,000 and 10,000) and a sliding window step 1 in order to get sufficient data to look at statistics of the $\chi^2$ divergence values. Tables~\ref{tab:non-malicioustab}, \ref{tab:malicioustab} and \ref{tab:mixturetab} show the results for the three different groups of datasets, respectively. Note that the $\chi^2$ divergence values of different datasets should not be compared directly as those datasets correspond to completely different networking environments and capturing methods. Despite the diversity of the datasets and the observable fluctuations of the $\chi^2$ divergence values especially the maximum values, largely speaking the $\chi^2$ divergence value does have a high tendency to following the following order: non-malicious $<$ mixed $<$ malicious. This suggests that the flow size difference may indeed be a good metric for building an anomaly-based IDS for detecting malicious TCP flows.

% Shujun: The data do not have consistent format (numbers of significant digits are different). Need alignment for final paper.

\begin{table*}[!htb]
\caption{$\chi^2$ divergence values for datasets with non-malicious TCP flows.}
\label{tab:non-malicioustab}
\centering\small
\begin{tabular}{r|r|cccc}
\hline
\multirow{2}{*}{Dataset (File)} & \multirow{2}{*}{Window-Size} & \multicolumn{4}{c}{$\chi^2$ divergence}\\
\cline{3-6}
& & Average & Median & Minimum & Maximum\\
\hline
LBNL/ICSI & 2000 & 0.06087285 & 0.02200721 & 0.000367677 & 3.72605\\
		  & 10000 & 0.023379069 & 0.01456664 & 0.000836337 & 0.14195\\
\hline
ISCX 2012 (11 June, 2010) & 2000 & 0.035543556 & 0.03086359 & 0.005292382 & 0.11853\\
    		              & 10000 & 0.024754012 & 0.02432506 & 0.014170007 & 0.03895\\
\hline
ISCX 2012 (12 June, 2010) & 2000 & 0.076106490 & 0.03924496 & 0.001458041 & 0.78477\\
            		      & 10000 & 0.043883460 & 0.03095162 & 0.003635401 & 0.18417 \\
\hline
TLERH & 2000 & 0.062321612 & 0.04681540 & 0.008825445 & 0.17574\\
      & 10000 & 0.047089776 & 0.04599299 & 0.036788053 & 0.05863\\
\hline
Twente & &\\
(loc1-20020523-1835) & 2000 & 0.010627572 & 0.00900225 & 0.001288395 & 0.03760\\
                     & 10000 & 0.005360776 & 0.00533056 & 0.002594503 & 0.00928\\
(loc1-20020524-1115) & 2000 & 0.008654660 & 0.00792808 & 0.001701887 & 0.02546\\
                     & 10000 & 0.003225103 & 0.00302386 & 0.001037643 & 0.00657\\
(loc2-20030513-1005) & 2000 & 0.017321427 & 0.01579575& 0.003077555 & 0.05431\\
                     & 10000 & 0.006945227 & 0.00694522 & 0.006945227 & 0.00694\\
(loc2-20030513-1044) & 2000 & 0.045903123 & 0.04062594 & 0.015966573 & 0.09972\\
                     & 10000 & 0.029621174 & 0.02962117 & 0.029621174 & 0.02962\\
(loc3-20030902-0930) & 2000 & 0.018619384 & 0.01775061 & 0.004399072 & 0.03852\\
                     & 10000 & 0.012487729 & 0.01280767 & 0.008713071 & 0.01677\\
(loc3-20030902-1005) & 2000 & 0.015866801 & 0.01588986 & 0.002744006 & 0.03715\\
                     & 10000 & 0.011495219 & 0.01157011 & 0.006588404 & 0.01691\\
(loc4-20040204-2145) & 2000 & 0.017711165 & 0.01771116 & 0.001162426 & 0.04583\\
                     & 10000 & 0.013173468 & 0.01434036 & 0.001360714 & 0.02149\\
(loc4-20040205-0410) & 2000 & 0.020032510 & 0.01927031 & 0.002402937 & 0.05800\\
                     & 10000 & 0.015180323 & 0.01636191 & 0.001294423 & 0.02562\\
(loc5-20031205-1431) & 2000 & 0.047492841 & 0.04569088 & 0.013472842 & 0.09680\\
                     & 10000 & 0.042015498 & 0.04209131 & 0.030077191 & 0.05689\\
(loc5-20031206-0731) & 2000 & 0.089616951 & 0.08408784& 0.048919967 & 0.18055\\
                     & 10000 & 0.070355597 & 0.07111595 & 0.061251126 & 0.07703\\
(loc6-20070501-2055) & 2000 & 0.060355909 & 0.04262247& 0.003198521 & 0.74357\\
                     & 10000 & 0.040565069 & 0.03629526 & 0.011316646 & 0.14548\\
(loc6-20070531-2043) & 2000 & 0.064672661 & 0.04806397& 0.003572854 & 0.90961\\
                     & 10000 & 0.048209326 & 0.03963367 & 0.009129598 & 0.19327\\
\hline
\end{tabular}
\end{table*}

\begin{table*}[!htb]
\caption{$\chi^2$ divergence values for datasets with malicious TCP flows.}
\label{tab:malicioustab}
\centering\small
\begin{tabular}{r|r|cccc}
\hline
\multirow{2}{*}{Dataset (File)} & \multirow{2}{*}{Window-Size} & \multicolumn{4}{c}{$\chi^2$ divergence}\\
\cline{3-6}
& & Average & Median & Minimum & Maximum\\
\hline
Capture Hacker & &\\
(HONEYBOT) & 2000 & 0.359898126 & 0.35989812 & 0.359898126 & 0.35989\\
(AMAZON) & 2000 & 0.541314000 & 0.66659157 & 0.291432604 & 0.77217\\
(NAPENTHES) & 2000 & 0.702143438 & 0.70214343& 0.702143438 & 0.70214\\ 		
\hline
2009 CDX & &\\
(dmp1) & 2000 & 0.416678762 & 0.39091630 & 0.155110896 & 0.70031\\
(dmp2) & 2000 & 0.531123191 & 0.45437578 & 0.235682400 & 1.03259\\
(dmp3) & 2000 & 1.290143507 & 0.54403769& 0.149578337 & 8.31071\\
(dmp4) & 2000 & 0.608195471 & 0.57431954 & 0.574319544 & 1.24340\\
(35dump*) & 2000 & 3.611008194 & 1.08928715 & 0.059356144 & 12.1582\\
(35dump2*) & 2000 & 0.531123191 & 0.45437578 & 0.235682400 & 1.03259\\
\hline
MACCDC & &\\
(MACC1) & 2000 & 0.561998014 & 0.49849047 & 0.030903985 & 2.48297820141\\
        & 10000 & 0.361823385 & 0.31002427 & 0.093006551 & 0.66489076768\\
(MACC2) & 2000 & 0.316137172 & 0.34970931 & 0.044942870 & 0.73696376246\\
        & 10000 & 0.302343885 & 0.33391531 & 0.077787425 & 0.48066011763\\  			
(MACC3) & 2000 & 0.479926028 & 0.51570687& 0.079583727 & 1.33988196727\\
        & 10000 & 0.279381711 & 0.27938171 & 0.279381711 & 0.27938171156\\
(MACC4) & 2000 & 0.452494598 & 0.41510598& 0.029984994 & 1.61375634415\\
        & 10000 & 0.317442530 & 0.24475004 & 0.047428552 & 0.91769942126\\
\hline
Twente & 2000 & 0.517489073 & 0.42960371 & 0.131377693 & 2.322259\\
       & 10000 & 0.354369539 & 0.37784575 & 0.134733907 & 0.58659\\
\hline
Kyoto & &\\
(20061101) & 2000 & 0.107162457 & 0.09849514 & 0.029682783 & 0.22778\\
(20061102) & 2000 & 1.069680688 & 1.00134293 & 0.100525961 & 3.163229\\
(20061103) & 2000 & 0.108852843 & 0.11355006& 0.058599317 & 0.19626716314\\
(20061104) & 2000 & 0.370117226 & 0.44502530& 0.071407197 & 0.61333268918\\
\hline
\end{tabular}
\end{table*}

\begin{table*}[!htb]
\caption{$\chi^2$ divergence values for datasets with mixed malicious and non-malicious TCP flows.}
\label{tab:mixturetab}
\centering\small
\begin{tabular}{r|r|cccc}
\hline
\multirow{2}{*}{Dataset (File)} & \multirow{2}{*}{Window-Size} & \multicolumn{4}{c}{$\chi^2$ divergence}\\
\cline{3-6}
& & Average & Median & Minimum & Maximum\\
\hline
ISCX 2012 & &\\
(13 June, 2010) & 2000 & 0.270556952 & 0.10817436 & 0.005356864 & 2.32225\\
                & 10000 & 0.194346177 & 0.08364604 & 0.022479027 & 2.24784\\
(14 June, 2015) & 2000 & 0.100901660 & 0.04324848 & 0.003802617 & 0.91445\\
                & 10000 & 0.063876498 & 0.03732686 & 0.010087834 & 0.37501\\
(15 June, 2015) & 2000 & 0.125542348 & 0.05355104 & 0.001430369 & 0.83598\\
                & 10000 & 0.105044000 & 0.04077830 & 0.005300968 & 0.58069\\
(17 June, 2015) & 2000 & 0.069305007 & 0.05964825 & 0.007398862 & 0.32598\\
                & 10000 & 0.046324304 & 0.04327538 & 0.016792197 & 0.09444\\
\hline
ISOT & 2000 & 0.051551610 & 0.02080945 & 0.000893305 & 1.52736\\
     & 10000 & 0.019348089 & 0.01382354 & 0.000874441 & 0.41050\\
\hline
\end{tabular}
\end{table*}

\section{IDS Performance Evaluation}
\label{sec:IDS}

In this section we provide details of our further experiments to evaluate the performance of a simple IDS built on top of the identified flow size difference metric. As we mentioned above, the IDS has only one metric and its structure is simple, so our goal is not to make it outperform other more complicated IDSs but to provide direct evidence that even with such a single metric the IDS has a reasonably good performance so it has the potential to be used \emph{together} with other known metrics and methods to design flow-based network IDSs with an even better overall performance. Due to many well-known and complicated issues around performance evaluation of IDSs \cite{Tavallaee2010IDSevaluation, SommerPaxson_IDS2010, dravsar2014similarity}, it is actually not trivial to compare performance of multiple IDSs fairly even if we want. Although we will not report comparative results with other IDSs, we chose to release our source code so that other IDS researchers can easily verify the results we reported and incorporate our work into other IDSs to conduct further performance analysis. Our source code was written in MATLAB and can be downloaded from \url{http://www.hooklee.com/Papers/Data/BenfordsLawIDS/source_code.7z}. Note that the source code is written to produce the performance evaluation results reported in this section, rather than as a ready-to-use IDS, but it is straightforward to translate our code into a deployable IDS or incorporate our code into an existing IDS written in any programming language.

\subsection{Datasets Used}

As pointed out by many other IDS researchers \cite{SommerPaxson_IDS2010, Tavallaee2010IDSevaluation, dravsar2014similarity}, there is a general lack of \emph{public labelled} datasets for evaluating and comparing performance of IDSs. The two mostly-used datasets in the past were the DARPA dataset \cite{DARPAdataset} released in 1999 (and updated in 2000 with some specific scenarios), and the KDD Cup 1999 dataset \cite{KDDCup1999} which was derived from the DARPA dataset. Both datasets have serious design flaws known to the IDS community for a long time e.g.\ those attacks presented in the dataset are very out-dated. Their use is in general discouraged and most researchers go for \emph{closed} datasets obtained from their own networks or from industrial collaborators, which are seldom made public thus making reproduction of published results impossible. There are some recent efforts of creating new public labelled IDS datasets \cite{Tavallaee_NSL-KDDDataSet_CISDA2009, sperottoIPOM2009, Fontugne_MAWILab_CoNEXT2010}, but none of them do not have their own drawbacks.

For our work here, we decided to use a closed dataset we call ``TRT dataset'' hereinafter, which was provided by Thales UK, Research and Technology (TRT) for this research. The TRT dataset contains 17 days of network traffic from a medium-sized network. In total there are 430,939 TCP flows. The dataset contains two subsets. The first subset was captured in a simulated network, with long periods of (simulated) normal traffic and occasional attacks of different types. The ``inet'' probe was located at the boundary of the network, to intercept all traffic entering and leaving the network from the Internet. The second subset was from a real student cyber security laboratory. It contains a lot of background traffic when the students were not there, but this was largely UDP traffic and hence not used in our experiments. When the students were there, they were doing a lot of (simulated) attacks as part of their practice and learning, and hence there is little ``normal'' TCP traffic in the data. Putting both subsets together, we have a balanced dataset with both periods with normal and malicious activities to test our IDS. The TRT dataset was originally not labelled but textual description about each day's activity was provided, and the labelling was done for this research by a volunteer (Florian Gottwalt) who was working at TRT while the research was taking place. Although we are unable to share the raw data of the TRT dataset, we decided to share the processed intermediate data (similarity metrics of all settings) at \url{http://www.hooklee.com/Papers/Data/BenfordsLawIDS/TRT_all_metrics.7z}, which can work with our source code to reproduce the experimental results reported in this section.

In addition to the above TRT dataset, to provide some form of reproducibility for  the experimental results reported in this paper, we also report the experimental results of applying our IDS to one public dataset: the KDD Cup 1999 dataset \cite{KDDCup1999}. This dataset has attacked connections and normal connections simulated in a military network environment. We used 1,870,598 TCP flows in KDD Cup 1999, which can be downloaded from \url{http://www.hooklee.com/Papers/Data/BenfordsLawIDS/kddcup_data_tcp.7z} in a (CSV) format ready to work with our source code. The KDD Cup 1999 dataset has a more recently-improved version, the NLD-KDD 2009 dataset \cite{NSL-KDDDataSet2009}, but we decided to stick with the original KDD dataset because 1) the NLD-KDD 2009 dataset was obtained by cleaning and resampling the original KDD dataset which destroys the original natural order of TCP flows; 2) the NLD-KDD 2009 dataset contains much less flows thus does not provide sufficient data for large window size; 3) our experiments showed NLD-KDD 2009 dataset is indeed not suitable for our experiments with very unstable results.

\subsection{Dataset Pre-Processing}

All the three datasets used have labels at the flow level, but our IDS works at the flow window level, so we needed to convert flow labels to flow window labels. The most natural way of doing the conversion is to introduce a threshold $T_l\in\{1,\cdots,W\}$ and then label a flow window as malicious if there are at least $T_l$ malicious flows. More precisely, assuming a flow window contains $W$ flows whose labels are $\{l_i=0\text{ or }1\}_{i=1}^W$ (1 means malicious and 0 means normal), then this flow window's label $L$ can be determined as follows:
\iffulledition
\begin{equation}\label{eq:FlowWindowLabel}
L = \text{sign}\left(\left(\sum_{i=1}^Wl_i\right)-T_l\right),
\end{equation}
where
\[
\text{sign}(x)=
\begin{cases}
1, & \text{if }x\geq0,\\
0, & \text{otherwise}.
\end{cases}
\]
\else
\begin{equation}\label{eq:FlowWindowLabel}
L = \text{sign}\left(\left(\sum\nolimits_{i=1}^Wl_i\right)-T_l\right).
\end{equation}
\fi
Note that the threshold $T_l$ can also be represented as a relative ratio between 0 and 1: $t_l=T_l/W$.

\subsection{Benchmarking Metric}

Instead of reporting detection accuracy indicators such as false positive and false negative rates, we decided to use the AUC (area under curve) of the ROC (receiver characteristic curve) of the threshold-based IDS as our main benchmarking metric. This is because our IDS has tunable parameters so reporting accuracy figures alone is too simplistic. We were aware that the ROC is not the best benchmarking tool for IDSs, but found it more straightforward to use for our proposed IDS to demonstrate the usefulness of the flow size difference as a metric for IDS purposes. Advanced benchmarking metrics normally introduce new factors (e.g.\ costs \cite{GaffneyUlvila2001IDSevaluation}) or additional steps (e.g.\ adaptive thresholding \cite{Ali2013ATT4IDS}), which we would like to avoid as they can unnecessarily over-complicate the performance evaluation task. Since we expect that the flow size difference as a metric will be used together with other metrics and methods to design more complicated IDSs, we leave investigation on advanced benchmarking metrics as our future work.

\subsection{Experimental Results}

We conducted extensive experiments on different parameter settings of the IDS with the TRT and the KDD Cup 1999 datasets. For each dataset, we consider full combinations of the following settings: zero handling (counting or skipping zeros), flow size definition (by byte or packet), relative or absolute threshold for labelling flow windows ($T_l$ or $t_l$), and all the seven similarity metrics we listed in Sec~\ref{sec:GoodnessMetrics}. The TRT dataset does not actually have any zero flow size difference and the KDD Cup dataset has only one definition of flow size (by byte), so these two options are irrelevant. The window sizes we tested include 100, 200, 500, 1000, 2500 and 5000. We tested 22 values of the relative threshold $t_l$ \iffulledition (0.01, 0.02, ..., 0.09, 0.1, 0.12, 0.14, 0.16, 0.18, 0.2, 0.3, ..., 0.9)\else from 0.01 and 0.9 \fi\ and 32 values of the absolute threshold $T_l$ \iffulledition (1, 2, ..., 9, 10, 20, 30, ..., 90, 100, 200, ..., 900, 1000, 2000, 3000, 4000, 5000)\else from 1 to 5000\fi.

\begin{figure*}[!htb]
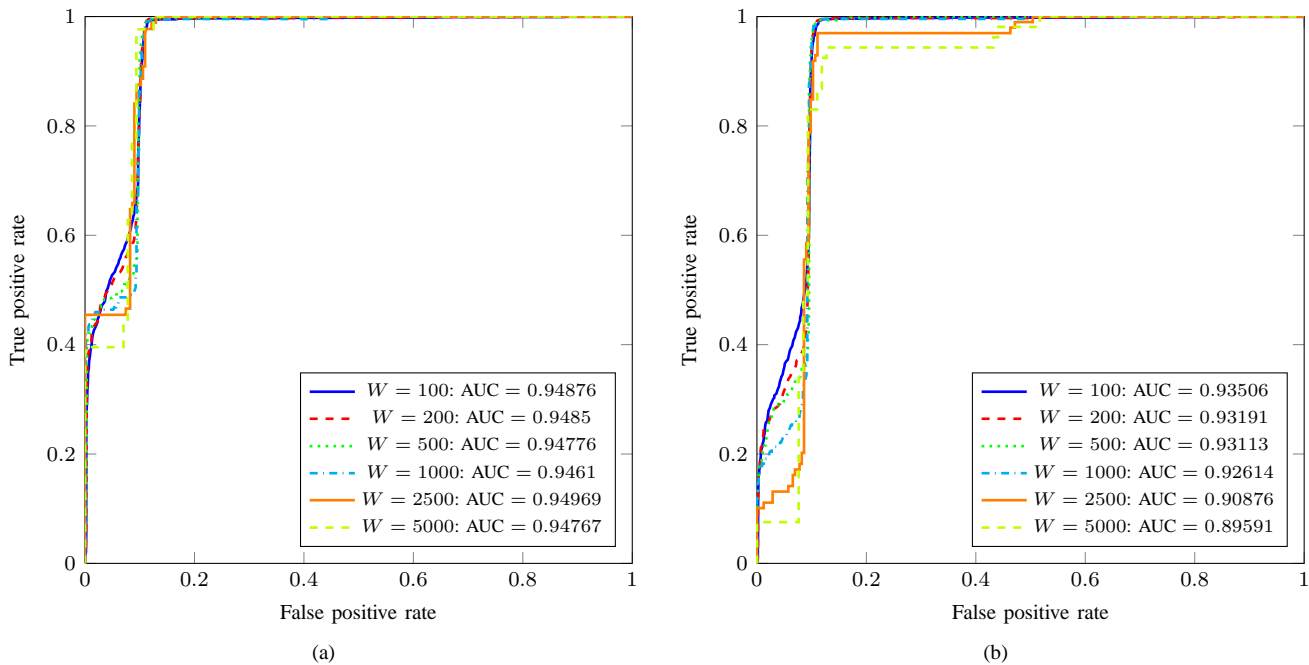

\subfloat[]{\input{fig/TRT_c2divs_tl0.4.tex}}
\quad
\subfloat[]{\input{fig/TRT_kldivs_Tl70.tex}}
\caption{ROC curves of the IDS tested with the TRT dataset: (a)  $\chi^2$ divergence, $t_l=0.4$; (b) modified Kullback-Leiber divergence, $T_l=70$.}\label{fig:IDS_TRT}
\end{figure*}

Largely speaking, the IDS worked very well for both datasets but it worked much better for the KDD Cup 1999 dataset which was expected because the TRT dataset contains more recent and more sophisticated attacks. For the KDD Cup 1999 dataset, for each setting we found at least one combination of the window size and the threshold to achieve a 0\% zero rate (i.e., the AUC being 1). Some selected ROC curves with the corresponding AUC values are shown in Figs.~\ref{fig:IDS_KDDcup1999} and \ref{fig:IDS_TRT}. A complete set of all AUC values with a comparative summary for all settings can be downloaded at \url{http://www.hooklee.com/Papers/Data/BenfordsLawIDS/AUCs_all.7z} (two Excel files, one dataset each).

\begin{figure*}[!htb]
\subfloat[]{\input{fig/KDDcup1999_c2divs_tl0.9.tex}}
\quad
\subfloat[]{\input{fig/KDDcup1999_kldivs_Tl1000.tex}}
\caption{ROC curves of the IDS tested with the KDD Cup 1999 dataset: (b) $\chi^2$ divergence, $t_l=0.9$; (b) modified Kullback-Leiber divergence, $T_l=1000$.}\label{fig:IDS_KDDcup1999}
\end{figure*}

All the similarity metrics worked reasonably well. The $\chi^2$ divergence and the modified Kullback-Leiber divergence performed the best overall cross both datasets and all settings.

For two different variants of the flow size difference (in byte and in packet) for the TRT dataset, the byte-based version worked consistently better: for the byte-based version the maximum AUC value across all settings is 0.949688057, and for the packet-based version the maximum value is only 0.926641651 ($\approx 2.43\%$ smaller).

The TRT dataset does not contain any zero flow size difference, so the zero handling option is irrelevant. As a contrast, the KDD Cup 1999 dataset has a large number of zero flow size differences, so both options of zero handling were tested. The experimental results revealed that it is more beneficial to count such zeros rather than skipping them.

The IDS's performance has dependency on the flow window size $W$ and the threshold $T_l$ ($t_l$), because $W$ can influence the temporal resolution and both parameters influence the definition of a flow window being considered malicious. Note that the threshold $T_l$ ($t_l$) is not an \emph{operational} parameter since it will not influence the output of the IDS, although it is needed to estimate detection accuracy. The detection threshold of the IDS is the real operational parameter which need frequently updating to balance the false negative and false positive rates.

\section{Discussions and Future Work}
\label{sec:discussion}

The results shown in the previous section should be seen with caution as for both tested datasets the attacks are not necessary representative in other networking environments. The ground-truth labels in the TRT dataset were manually done by a human expert who looked at the network traffic data without information of what actually happened during the traffic recording sessions. To some extent we can consider the TRT dataset's labels as ratings of a human expert on suspiciousness of flows, so putting all the results together, we can argue that the proposed IDS does work in predicting what a human expert would produce. The fact that the proposed IDS works on both a public dataset and a more recent and independent dataset implies that the proposed method is robust as well. In future we plan to test the proposed IDS on more datasets to further verify its performance across more real-world datasets. We also plan to make use of the automatically labelled (using an ensemble of automated IDSs) MAWILab dataset \cite{Fontugne_MAWILab_CoNEXT2010} to test our work against other automated IDSs on a more diverse set of network traffic.

The flow size difference as a new IDS metric can obviously be used to enhance other existing IDSs detecting malicious network flows. Some possible ways to achieve this include: 1) adding the flow size difference as a new feature used by the IDS classifier together with other features; 2) using the flow size difference based IDS as an early warning system to trigger other more advanced IDSs to focus on regions of interests in the network traffic; 3) using the flow size difference based IDS as a secondary IDS to identify missing attacks; 4) using the flow size difference based IDS to generate pseudo-labels for semi-supervised training of a more complicated IDS.

Using the simple form of Benford's law without any generalization means no parameter estimation is needed. In addition, the flow size difference based IDS is a simple threshold based IDS without the need to be trained (which is a common merit of all distribution based IDSs). The IDS does have some adjustable parameters but they can be set intuitively and based on observed statistics of the similarity metric, which can be done by a simple software tool processing all historical flows (both normal and malicious ones). Even when such historical data is not available, one can still set some initial values and then incrementally adjust the parameters while normal and malicious flows are being observed. As a whole, the IDS can be easily deployed without much burden of ``making it ready''.

The two main operational parameters of the flow size difference based IDS are: the classifier's decision threshold $T$, the flow window size $W$. The former can be used to balance false negative and false positive rates, and the latter to balance detection accuracy and temporal resolution. Note that $W$ will influence the false negative and false positive rates as well (as shown in the previous section) but in an indirect way. Both parameters have a clear semantic meaning from an IDS operator's point of view: 1) $T$ is the deviation from Benford's law, as measured by the selected similarity metric; 2) $W$ simply define how many flows one need to observe before making a decision. Typical initial values of both parameters can be set to $T=0.4$ and $W=2500$ according to our experiments on the TRT dataset. We believe that $W$ is more stable so once a preferred value is set it will remain for a long time without the need to be changed, but $T$ may evolve more quickly in order to adapt to rapidly evolving attacks.

If an attacker is aware of the use of the flow size difference based IDS, he will try to adapt his attacking strategies so that the network traffic generated from attacks can fall below the alerting threshold. There are at least two possible strategies: 1) adapting the attacks' network traffic precisely to match the target distribution (i.e., Benford's law), which is difficult or impossible as the attacker normally does not have control of other users and network devices' activities; 2) reducing the level of attack activity so that it hides well in normal traffic, which effectively leads to a special kind of network-based information hiding (which will require a completely different treatment for detection). The second strategy may make some attacks completely meaningless (harmless) e.g.\ for DDoS slowing down the incoming traffic will make the attack itself fail by definition. For some other attacks, the second strategy will normally be acceptable (e.g.\ port scanning or online password guess) but will significantly increase the costs of performing the target attacks and the chance of the attacks being detected by other means (e.g.\ other IDS methods or manual inspection). If launching an attack requires too much time, the target may also become unavailable e.g.\ if the target is a moving object or does not have a permanent network address -- a typical example is a mobile device.

We focused at bidirectional TCP flows in this work. Our initial tests on other types of network flows such as unidirectional TCP flows showed that Benford's law may not be applied directly, which can be explained by the observation that many unidirectional TCP flows are often highly asymmetric (more and larger packets from server to client). Possible applications to other types of flows such as IP, ICMP and UDP flows require more work as well since such flows are normally defined as unidirectional and the sessional boundary is not as clearly defined as in the TCP case.

In our future work we will also investigate if and how our work can be generalized beyond the simple form of Benford's law. For instance, we will look at Benford's law with a base different from 10, the generalized Benford's law \cite{Fu} and other natural laws such as Zipf's law \cite{newman2005power, Tao2009ThreeLaws} and the Stigler distribution \cite{Lee2010StiglerLaw} (in case the simple Benford's law does not work in some cases e.g.\ UDP flows).

\section{Conclusion}

This paper investigates possible applications of Benford's law to anomaly-based network flow IDS and reports a new metric which has not been studied: flow size difference. Different from flow size, flow size difference has been rarely studied in the IDS and networking literature, but our study showed that it is a very good metric for IDS purposes: normal TCP flows follow Benford's law closely but malicious ones deviate from it significantly. We conducted a large number of experiments with many network traffic datasets to verify the observation, and used it to construct a simple threshold-based IDS. The IDS was tested with two labelled datasets (one closed dataset and one public dataset) and it has been shown the performance of the IDS is promising, especially considering its simple structure. We call for more research on the use of flow size difference in IDS research, particularly on how to use it to improve other IDS methods.

\section*{Acknowledgments}

Aamo Iorliam would like to acknowledge the financial support of the Benue State University, Nigeria for his PhD scholarship. The authors would also like to acknowledge Florian Gottwalt for labeling the TRT dataset.

\iffulledition

\section*{Contributions}

Anthony T.S.\ Ho and Adrian Waller proposed the original idea of applying Benford's law to network traffic analysis, and all co-authors participated in designs of the experiments and the IDS reported in this paper. Aamo Iorliam and Santosh Tirunagari wrote the source code related to Sec.~\ref{sec:Experiments} and conducted all experiments. Shujun Li, Santosh Tirunagari and Aamo Iorliam wrote the source code related to Sec.~\ref{sec:IDS}. Shujun Li, Aamo Iorliam and Santosh Tirunagari conducted the experiments reported in Sec.~\ref{sec:IDS}, and Adrian Waller provided support for experiments on the TRT dataset. Aamo Iorliam was the principal writer of a first draft of the paper and a chapter of his thesis, and Shujun Li was the principal writer of the final draft of the paper. All co-authors contributed to writing of the paper.

\fi

\ifCLASSOPTIONcaptionsoff
  \newpage
\fi


\bibliographystyle{IEEEtran}
\iffulledition
  \bibliography{network_traffic_full}
\else
  \bibliography{network_traffic}

\begin{thebibliography}{10}
\providecommand{\url}[1]{#1}
\csname url@samestyle\endcsname
\providecommand{\newblock}{\relax}
\providecommand{\bibinfo}[2]{#2}
\providecommand{\BIBentrySTDinterwordspacing}{\spaceskip=0pt\relax}
\providecommand{\BIBentryALTinterwordstretchfactor}{4}
\providecommand{\BIBentryALTinterwordspacing}{\spaceskip=\fontdimen2\font plus
\BIBentryALTinterwordstretchfactor\fontdimen3\font minus
  \fontdimen4\font\relax}
\providecommand{\BIBforeignlanguage}[2]{{%
\expandafter\ifx\csname l@#1\endcsname\relax
\typeout{** WARNING: IEEEtran.bst: No hyphenation pattern has been}%
\typeout{** loaded for the language `#1'. Using the pattern for}%
\typeout{** the default language instead.}%
\else
\language=\csname l@#1\endcsname
\fi
#2}}
\providecommand{\BIBdecl}{\relax}
\BIBdecl

\bibitem{DDoS_Survey2013}
S.~T. Zargar, J.~Joshi, and D.~Tipper, ``A survey of defense mechanisms against
  distributed denial of service ({DDoS}) flooding attacks,'' \emph{IEEE
  Communications Surveys \& Tutorials}, vol.~15, no.~4, pp. 2046--2069, 2013.

\bibitem{BotnetBook2008}
W.~Lee, C.~Wang, and D.~Dagon, Eds., \emph{Botnet Detection: Countering the
  Largest Security Threat}, ser. Advances in Information Security.\hskip 1em
  plus 0.5em minus 0.4em\relax Springer-Verlag US, 2008, vol.~36.

\bibitem{Worm_Bookchapter2006}
J.~A. Cock, ``Worms,'' in \emph{Computer Viruses and Malware}.\hskip 1em plus
  0.5em minus 0.4em\relax Springer Science+Business Media, 2006, ch.~7, pp.
  143--155.

\bibitem{PhishingBook2006}
M.~Jakobsson and S.~Myers, Eds., \emph{Phishing and Countermeasures:
  Understanding the Increasing Problem of Electronic Identity Theft}.\hskip 1em
  plus 0.5em minus 0.4em\relax John Wiley \& Sons, Inc., 2006.

\bibitem{DDoS_BBC}
C.~Baraniuk, ``{DDoS}: Website-crippling cyber-attacks to rise in 2016,'' BBC
  news, \url{http://www.bbc.co.uk/news/technology-35376327}, 2016.

\bibitem{DDoS_500Gbps}
D.~Pauli, ``500{G}bps {DDoS} attack flattens world record,'' The Register,
  \url{http://www.theregister.co.uk/2016/01/27/500gbps_ddos_attack_flattens_world_record/},
  2016.

\bibitem{kruegel2005intrusion}
C.~Kruegel, F.~Valeur, and G.~Vigna, \emph{Intrusion Detection and Correlation:
  Challenges and Solutions}, ser. Advances in Information Security.\hskip 1em
  plus 0.5em minus 0.4em\relax Springer Science + Business Media, Inc., 2005,
  vol.~14.

\bibitem{NIDS_Book2009}
A.~A. Ghorbani, W.~Lu, and M.~Tavallaee, Eds., \emph{Network Intrusion
  Detection and Prevention: Concepts and Techniques}.\hskip 1em plus 0.5em
  minus 0.4em\relax Springer Science+Business Media, 2009.

\bibitem{Debar_IDS_taxonomy2000}
H.~Debar, M.~Dacier, and A.~Wespi, ``A revised taxonomy for intrusion-detection
  systems,'' \emph{Annales Des T\'{e}l\'{e}communications}, vol.~55, no.~7, p.
  361–378, 2000.

\bibitem{dravsar2014similarity}
M.~Dra{\v{s}}ar, M.~Vizv{\'a}ry, and J.~Vykopal, ``Similarity as a central
  approach to flow-based anomaly detection,'' \emph{International Journal of
  Network Management}, vol.~24, no.~4, pp. 318--336, 2014.

\bibitem{dreger2004operational}
H.~Dreger, A.~Feldmann, V.~Paxson, and R.~Sommer, ``Operational experiences
  with high-volume network intrusion detection,'' in \emph{Proceedings of 11th
  ACM conference on Computer and Communications Security}.\hskip 1em plus 0.5em
  minus 0.4em\relax ACM, 2004, pp. 2--11.

\bibitem{bejtlich2004tao}
R.~Bejtlich, \emph{The Tao of Network Security Monitoring: Beyond Intrusion
  Detection}.\hskip 1em plus 0.5em minus 0.4em\relax Addison Wesley, 2004.

\bibitem{sperotto2011flow}
A.~Sperotto and A.~Pras, ``Flow-based intrusion detection,'' in
  \emph{Proceedings of 2011 IFIP/IEEE International Symposium on Integrated
  Network Management (IM)}.\hskip 1em plus 0.5em minus 0.4em\relax IEEE, 2011,
  pp. 958--963.

\bibitem{steinberger2013anomaly}
J.~Steinberger, L.~Schehlmann, S.~Abt, and H.~Baier, ``Anomaly detection and
  mitigation at {Internet} scale: A survey,'' in \emph{Emerging Management
  Mechanisms for the Future Internet: 7th IFIP WG 6.6 International Conference
  on Autonomous Infrastructure, Management, and Security, AIMS 2013, Barcelona,
  Spain, June 25-28, 2013. Proceedings}, ser. Lecture Notes in Computer
  Science, vol. 7943.\hskip 1em plus 0.5em minus 0.4em\relax Springer Berlin
  Heidelberg, 2013, pp. 49--60.

\bibitem{newman2005power}
M.~E.~J. Newman, ``Power laws, {Pareto} distributions and {Zipf's} law,''
  \emph{Contemporary Physics}, vol.~46, no.~5, pp. 323--351, 2005.

\bibitem{Tao2009ThreeLaws}
T.~Tao, ``{Benford}'s law, {Zipf}'s law, and the {Pareto} distribution,''
  \url{https://terrytao.wordpress.com/2009/07/03/benfords-law-zipfs-law-and-the-pareto-distribution/},
  2009.

\bibitem{BenfordsLawBook2015}
A.~Berger and T.~P. Hill, \emph{An Introduction to {Benford}'s Law}.\hskip 1em
  plus 0.5em minus 0.4em\relax Princeton University Press, 2015.

\bibitem{WeibullDistribution}
Y.~K. Belyaev and E.~V. Chepurin, ``{Weibull} distribution,'' in
  \emph{Encyclopedia of Mathematics}.\hskip 1em plus 0.5em minus 0.4em\relax
  Springer and European Mathematical Society,
  \url{http://www.encyclopediaofmath.org/index.php?title=Weibull_distribution&oldid=18906}.

\bibitem{yegneswaran2003internet}
V.~Yegneswaran, P.~Barford, and J.~Ullrich, ``Internet intrusions: Global
  characteristics and prevalence,'' \emph{ACM SIGMETRICS Performance Evaluation
  Review}, vol.~31, no.~1, pp. 138--147, 2003, proceedings of 2003 ACM
  SIGMETRICS International Conference on Measurement and Modeling of Computer
  Systems (SIGMETRICS '03).

\bibitem{Hamadeh_MScThesis2004}
N.~Hamadeh, ``Wireless security and traffic modeling using {Benford}'s law,''
  Master's Thesis, University of New Mexico, USA, 2004.

\bibitem{Wang2004}
K.~Wang and S.~J. Stolfo, \emph{Anomalous Payload-Based Network Intrusion
  Detection}, ser. Lecture Notes in Computer Science.\hskip 1em plus 0.5em
  minus 0.4em\relax Springer Berlin Heidelberg, 2004, vol. 3224, pp. 203--222.

\bibitem{Cai2007}
M.~Cai, K.~Hwang, J.~Pan, and C.~Papadopoulos, ``{WormShield}: Fast worm
  signature generation with distributed fingerprint aggregation,'' \emph{IEEE
  Transactions on Dependable and Secure Computing}, vol.~4, no.~2, pp. 88--104,
  2007.

\bibitem{Gu_USENIX_Security2007}
G.~Gu, P.~Porras, V.~Yegneswaran, M.~Fong, and W.~Lee, ``{BotHunter}: Detecting
  malware infection through {IDS}-driven dialog correlation,'' in
  \emph{Proceedings of 16th USENIX Security Symposium}.\hskip 1em plus 0.5em
  minus 0.4em\relax USENIX Association, 2007, pp. 167--182.

\bibitem{Kim_AINAW2007}
B.~Kim and S.~Bahk, ``Relative entropy-based filtering of internet worms by
  inspecting {TCP SYN} retry packets,'' in \emph{Proceedings of 21st
  International Conference on Advanced Information Networking and Applications
  Workshops (AINAW'07)}.\hskip 1em plus 0.5em minus 0.4em\relax IEEE, 2007.

\bibitem{YeZheng2011}
C.~Ye and K.~Zheng, ``Detection of application layer distributed denial of
  service,'' in \emph{Proceedings of 2011 International Conference on Computer
  Science and Network Technology}.\hskip 1em plus 0.5em minus 0.4em\relax IEEE,
  2011, pp. 310--314.

\bibitem{Arshadi_WeibullIDS_EMS2011}
L.~Arshadi and A.~H. Jahangir, ``On the {TCP} flow inter-arrival times
  distribution,'' in \emph{Proceedings of 2011 UKSim 5th European Symposium on
  Computer Modeling and Simulation (EMS 2011)}.\hskip 1em plus 0.5em minus
  0.4em\relax IEEE, 2011, pp. 360--365.

\bibitem{yu2012discriminating}
S.~Yu, W.~Zhou, W.~Jia, S.~Guo, Y.~Xiang, and F.~Tang, ``Discriminating {DDoS}
  attacks from flash crowds using flow correlation coefficient,'' \emph{IEEE
  Transactions on Parallel and Distributed Systems}, vol.~23, no.~6, pp.
  1073--1080, 2012.

\bibitem{arshadi2014empirical}
L.~Arshadi and A.~H. Jahangir, ``An empirical study on {TCP} flow interarrival
  time distribution for normal and anomalous traffic,'' \emph{International
  Journal of Communication Systems}, 2014.

\bibitem{Newcomb1881}
S.~Newcomb, ``Note on the frequency of use of the different digits in natural
  numbers,'' \emph{American Journal of Mathematics}, vol.~4, no.~1, p. 39–40,
  1881.

\bibitem{Benford1938}
F.~Benford, ``The law of anomalous numbers,'' \emph{Proceedings of the American
  Philosophical Society}, vol.~78, no.~4, pp. 551--572, 1938.

\bibitem{Formann2010BenfordLaw}
A.~K. Formann, ``The {Newcomb-Benford} law in its relation to some common
  distributions,'' \emph{PLoS One}, vol.~5, no.~5, p. e10541, 2010.

\bibitem{BergerHill2011BenfordSurvey}
A.~Berger and T.~P. Hill, ``A basic theory of {Benford}'s law,''
  \emph{Probability Surveys}, vol.~8, pp. 1--126, 2011.

\bibitem{Pinkham1961BenfordLaw}
R.~S. Pinkham, ``On the distribution of first significant digits,''
  \emph{Annals of Mathematical Statistics}, vol.~32, no.~4, pp. 1223--1230,
  1961.

\bibitem{Hill1995BenfordLaw}
T.~P. Hill, ``Base-invariance implies {Benford}'s law,'' \emph{Proceedings of
  the American Mathematical Society}, vol. 123, no.~3, pp. 887--895, 1995.

\bibitem{Hill1995}
------, ``A statistical derivation of the significant-digit law,''
  \emph{Statistical Science}, vol.~10, pp. 354--363, 1995.

\bibitem{BergerHill2011BenfordLaw}
A.~Berger and T.~P. Hill, ``{Benford}'s law strikes back: No simple explanation
  in sight for mathematical gem,'' \emph{The Mathematical Intelligencer},
  vol.~33, no.~1, pp. 85--91, 2011.

\bibitem{Lee2010StiglerLaw}
J.~Lee, W.~K.~T. Cho, and G.~G. Judge, ``{Stigler}'s approach to recovering the
  distribution of first significant digits in natural data sets,''
  \emph{Statistics \& Probability Letters}, vol.~80, no.~2, pp. 82--88, 2010.

\bibitem{irmay1997relationship}
S.~Irmay, ``The relationship between {Zipf}'s law and the distribution of first
  digits,'' \emph{Journal of Applied Statistics}, vol.~24, no.~4, pp. 383--394,
  1997.

\bibitem{Fu}
D.~Fu, Y.~Q. Shi, and W.~Su, ``A generalized {Benford}'s law for {JPEG}
  coefficients and its applications in image forensics,'' in \emph{Security,
  Steganography, and Watermarking of Multimedia Contents IX}, ser. Proceedings
  of SPIE, vol. 6505, 2007, p. 65051L.

\bibitem{BeebeBenfordBibTeX}
N.~H.~F. Beebe, ``{BibTeX} bibliography benfords-law.bib,'' Online document,
  Version 1.99, \url{http://ftp.math.utah.edu/pub/tex/bib/benfords-law.html},
  2016.

\bibitem{benfordonline}
A.~Berger, T.~Hill, and E.~Rogers, ``{Benford} online bibliography,'' Website,
  \url{http://www.benfordonline.net/}, 2009.

\bibitem{Durtschi2004BenfordLaw4FraudDetection}
C.~Durtschi, W.~Hillison, and C.~Pacini, ``The effective use of {Benford}'s law
  to assist in detecting fraud in accounting data,'' \emph{Journal of Forensic
  Accounting}, vol.~5, pp. 17--34, 2004.

\bibitem{DiekmannJann2010BenfordLawFraudDetection}
A.~Diekmann and B.~Jann, ``{Benford}'s law and fraud detection: Facts and
  legends,'' \emph{German Economic Review}, vol.~11, no.~3, p. 397–401, 2010.

\bibitem{Nigrini2012BenfordLawBook}
M.~Nigrini, \emph{{Benford}'s Law: Applications for Forensic Accounting,
  Auditing, and Fraud Detection}.\hskip 1em plus 0.5em minus 0.4em\relax John
  Wiley \& Sons, Inc., 2012.

\bibitem{Acebo}
E.~Acebo and M.~Sbert, ``{Benford}'s law for natural and synthetic images,'' in
  \emph{Proceedings of 1st Eurographics Conference on Computational Aesthetics
  in Graphics, Visualization and Imaging}, 2005, pp. 169--176.

\bibitem{li2008detecting}
B.~Li, Y.~Q. Shi, and J.~Huang, ``Detecting doubly compressed {JPEG} images by
  using mode based first digit features,'' in \emph{Proceedings of 2008 IEEE
  10th Workshop on Multimedia Signal Processing}.\hskip 1em plus 0.5em minus
  0.4em\relax IEEE, 2008, pp. 730--735.

\bibitem{li2012detection}
X.~H. Li, Y.~Q. Zhao, M.~Liao, F.~Y. Shih, and Y.~Q. Shi, ``Detection of
  tampered region for {JPEG} images by using mode-based first digit features,''
  \emph{EURASIP Journal on Advances in Signal Processing}, vol. 2012, no.~1,
  pp. 1--10, 2012.

\bibitem{Qadir}
G.~Qadir, X.~Zhao, and A.~T.~S. Ho, ``Estimating {JPEG2000} compression for
  image forensics using {Benford}'s law,'' in \emph{Optics, Photonics, and
  Digital Technologies for Multimedia Applications}, ser. Proceedings of SPIE,
  vol. 7723, 2010, p. 77230J.

\bibitem{Helman_IDS_Foundations_CSFW92}
P.~Helman, G.~Liepins, and W.~Richards, ``Foundations of intrusion detection,''
  in \emph{Proceedings of 1992 Computer Security Foundations Workshop V}.\hskip
  1em plus 0.5em minus 0.4em\relax IEEE, 1992, pp. 114--120.

\bibitem{Axelsson_BaseRateFallacy_CCS99}
S.~Axelsson, ``The base-rate fallacy and its implications for the difficulty of
  intrusion detection,'' in \emph{Proceedings of 6th ACM Conference on Computer
  and Communications Security (CCS'99)}.\hskip 1em plus 0.5em minus 0.4em\relax
  ACM, 1999, pp. 1--7.

\bibitem{SommerPaxson_IDS2010}
R.~Sommer and V.~Paxson, ``Outside the closed world: On using machine learning
  for network intrusion detection,'' in \emph{Proceedings of 2010 IEEE
  Symposium on Security and Privacy}.\hskip 1em plus 0.5em minus 0.4em\relax
  IEEE, 2010, pp. 205--316.

\bibitem{DenningIDS1987}
D.~E. Denning, ``An intrusion-detection model,'' \emph{IEEE Transactions on
  Software Engineering}, vol.~13, no.~2, pp. 222--232, 1987, an earlier edition
  of the paper was presented at 1986 IEEE Symposium of Security and Privacy.

\bibitem{arshadi2014benford}
L.~Arshadi and A.~H. Jahangir, ``{Benford}'s law behavior of {Internet}
  traffic,'' \emph{Journal of Network and Computer Applications}, vol.~40, pp.
  194--205, 2014.

\bibitem{Fredj2001flow}
S.~B. Fredj, T.~Bonald, A.~Proutiere, G.~R\'{e}gni\'{e}, and J.~W. Roberts,
  ``Statistical bandwidth sharing: A study of congestion at flow level,'' in
  \emph{Proceedings of 2001 Conference on Applications, Technologies,
  Architectures, and Protocols for Computer Communications (SIGCOMM
  2001)}.\hskip 1em plus 0.5em minus 0.4em\relax ACM, 2001, pp. 111--122.

\bibitem{tune2013internet}
\BIBentryALTinterwordspacing
P.~Tune and M.~Roughan, ``Internet traffic matrices: A primer,'' in
  \emph{Recent Advances in Networking}.\hskip 1em plus 0.5em minus 0.4em\relax
  ACM, 2013. [Online]. Available:
  \url{\url{http://sigcomm.org/education/ebook/SIGCOMMeBook2013v1_chapter3.pdf}}
\BIBentrySTDinterwordspacing

\bibitem{Hofstede2014flow}
R.~Hofstede, P.~\v{C}eleda, B.~Trammell, I.~Drago, R.~Sadre, A.~Sperotto, and
  A.~Pras, ``Flow monitoring explained: From packet capture to data analysis
  with {NetFlow} and {IPFIX},'' \emph{IEEE Communications Surveys \&
  Tutorials}, vol.~16, no.~4, pp. 2037--2064, 2014.

\bibitem{sperotto2010overview}
A.~Sperotto, G.~Schaffrath, R.~Sadre, C.~Morariu, A.~Pras, and B.~Stiller, ``An
  overview of {IP} flow-based intrusion detection,'' \emph{IEEE Communications
  Surveys \& Tutorials}, vol.~12, no.~3, pp. 343--356, 2010.

\bibitem{NFDUMP}
P.~Haag, ``{NFDUMP},'' \url{http://nfdump.sourceforge.net/}, 2014.

\bibitem{NfSen}
------, ``{NfSen - Netflow Sensor},'' \url{http://nfsen.sourceforge.net/},
  2014.

\bibitem{shiravi2012toward}
A.~Shiravi, H.~Shiravi, M.~Tavallaee, and A.~A. Ghorbani, ``Toward developing a
  systematic approach to generate benchmark datasets for intrusion detection,''
  \emph{Computers \& Security}, vol.~31, no.~3, pp. 357--374, 2012.

\bibitem{Kumar2004FSDestimate}
A.~Kumar, M.~Sung, J.~J. Xu, and J.~Wang, ``Data streaming algorithms for
  efficient and accurate estimation of flow size distribution,'' in
  \emph{Proceedings of 2004 Joint International Conference on Measurement and
  Modeling of Computer Systems (ACM SIGMETRICS '04/IFIP Performance
  '04)}.\hskip 1em plus 0.5em minus 0.4em\relax ACM, 2004, pp. 177--188.

\bibitem{Tune2014OFSS}
P.~Tune and D.~Veitch, ``{OFSS}: Skampling for the flow size distribution,'' in
  \emph{Proceedings of 2014 Conference on Internet Measurement Conference (IMC
  2014)}.\hskip 1em plus 0.5em minus 0.4em\relax ACM, 2014, pp. 235--240.

\bibitem{Garsva2015FlowSizeDistribution}
E.~Gar\v{s}va, N.~Paulauskas, and G.~Gra\v{z}ulevi\v{c}ius, ``Packet size
  distribution tendencies in computer network flows,'' in \emph{Proceedings of
  2015 Open Conference of Electrical, Electronic and Information Sciences
  (eStream 2015)}.\hskip 1em plus 0.5em minus 0.4em\relax IEEE, 2015.

\bibitem{Nychis2008entropyIDS}
G.~Nychis, V.~Sekar, D.~G. Andersen, H.~Kim, and H.~Zhang, ``An empirical
  evaluation of entropy-based traffic anomaly detection,'' in \emph{Proceedings
  of 2008 Conference on Internet Measurement Conference (IMC 2008)}.\hskip 1em
  plus 0.5em minus 0.4em\relax ACM, 2008, pp. 151--156.

\bibitem{Basicevic2016FSD_IDS}
I.~B. abd Stanislav~Ocovaj and M.~Popovic, ``The value of flow size
  distribution in entropy-based detection of {DoS} attacks,'' \emph{Security
  and Communication Networks}, vol.~9, no.~10, p. 958–965, 2016.

\bibitem{Plackett1983chi2test}
R.~L. Plackett, ``{Karl Pearson} and the chi-squared test,''
  \emph{International Statistical Review}, vol.~51, no.~1, pp. 59--72, 1983.

\bibitem{weller2015survey}
D.~J. Weller-Fahy, B.~J. Borghetti, and A.~A. Sodemann, ``A survey of distance
  and similarity measures used within network intrusion anomaly detection,''
  \emph{IEEE Communications Surveys \& Tutorials}, vol.~17, no.~1, pp. 70--91,
  2015.

\bibitem{LBNLdata}
{Lawrence Berkeley National Laboratory and ICSI}, ``{LBNL/ICSI} enterprise
  tracing project,'' \url{http://www.icir.org/enterprise-tracing}, 2005.

\bibitem{ISCX2012dataset}
{Information Security Center of eXcellence, University of New Brunswick, USA},
  ``{UNB ISCX} 2012 intrusion detection evaluation data set,''
  \url{http://www.unb.ca/research/iscx/dataset/iscx-IDS-dataset.html}, 2012.

\bibitem{szabo2010traffic}
G.~Szab{\'o}, I.~G{\'o}dor, A.~Veres, S.~Malomsoky, and S.~Moln{\'a}r,
  ``Traffic classification over {Gbit} speed with commodity hardware,''
  \emph{Journal of Communications Software and Systems}, vol.~5, no.~3, 2010.

\bibitem{TwenteIPOM2009Dataset}
R.~van Rijswijk-Deij, A.~Sperotto, and A.~Pras, ``{netflow2},''
  \url{https://traces.simpleweb.org/traces/netflow/netflow2/}, 2015.

\bibitem{sperottoIPOM2009}
A.~Sperotto, R.~Sadre, F.~{van~Vliet}, and A.~Pras, ``A labeled data set for
  flow-based intrusion detection,'' in \emph{IP Operations and Management: 9th
  IEEE International Workshop, IPOM 2009, Venice, Italy, October 29-30, 2009.
  Proceedings}, ser. Lecture Notes in Computer Science, vol. 5843.\hskip 1em
  plus 0.5em minus 0.4em\relax Springer Berlin Heidelberg, 2009, pp. 39--50.

\bibitem{CtH2013NAPENTHES}
D.~Day, ``Capture the hacker 2013 competition: {NAPENTHES} dataset,''
  \url{http://www.snaketrap.co.uk/pcaps/Ncapture.pcap}.

\bibitem{CtH2013HONEYBOT}
------, ``Capture the hacker 2013 competition: {HONEYBOT} dataset,''
  \url{http://www.snaketrap.co.uk/pcaps/hbot.pcap}.

\bibitem{CtH2013AMAZON}
------, ``Capture the hacker 2013 competition: {AMAZON} dataset,''
  \url{http://www.snaketrap.co.uk/pcap/hptcp.pcap}.

\bibitem{CDX_2009}
{Information Technology Operations Center}, ``Inter-service academy cyber
  defense competition,'' \url{https://www.itoc.usma.edu/research/dataset/},
  2009.

\bibitem{MACCDC_netresec}
{NETRESEC AB}, ``Capture files from {Mid-Atlantic CCDC},''
  \url{http://www.netresec.com/?page=MACCDC}, 2015.

\bibitem{KyotoHoneypotsDataset}
J.~Song, H.~Takakura, and Y.~Okabe, ``Traffic data from {Kyoto University}'s
  honeypots,'' \url{http://www.takakura.com/kyoto_data/}.

\bibitem{song2008cooperation}
------, ``Cooperation of intelligent honeypots to detect unknown malicious
  codes,'' in \emph{Proceedings of 2008 WOMBAT Workshop on Information Security
  Threats Data Collection and Sharing (WISTDCS 2008)}, 2008, pp. 31--39.

\bibitem{ISOTdataset}
{ISOT Research Lab}, ``Botnet datasets,''
  \url{http://www.uvic.ca/engineering/ece/isot/datasets/}.

\bibitem{saad2011detecting}
S.~Saad, I.~Traore, A.~Ghorbani, B.~Sayed, D.~Zhao, W.~Lu, J.~Felix, and
  P.~Hakimian, ``Detecting {P2P} botnets through network behavior analysis and
  machine learning,'' in \emph{Proceedings of 2011 9th Annual International
  Conference on Privacy, Security and Trust (PST 2011)}, 2011, pp. 174--180.

\bibitem{Tavallaee2010IDSevaluation}
M.~Tavallaee, N.~Stakhanova, and A.~A. Ghorbani, ``Toward credible evaluation
  of anomaly-based intrusion-detection methods,'' \emph{IEEE Transactions on
  Systems, Man, and Cybernetics, Part C: Applications and Reviews}, vol.~40,
  no.~5, pp. 516--524, 2010.

\bibitem{DARPAdataset}
{Cyber Systems and Technology Group, MIT Lincoln Laboratory}, ``{DARPA}
  intrusion detection data sets,'' \url{http://www.ll.mit.edu/ideval/data/},
  2000.

\bibitem{KDDCup1999}
{Information and Computer Science, University of California, Irvine}, ``{KDD
  Cup} 1999 data,''
  \url{http://kdd.ics.uci.edu/databases/kddcup99/kddcup99.html}, 1999.

\bibitem{Tavallaee_NSL-KDDDataSet_CISDA2009}
M.~Tavallaee, E.~Bagheri, W.~Lu, and A.~A. Ghorbani, ``A detailed analysis of
  the {KDD CUP} 99 data set,'' in \emph{Proceedings of 2009 IEEE Symposium on
  Computational Intelligence in Security and Defense Applications (CISDA
  2009)}.\hskip 1em plus 0.5em minus 0.4em\relax IEEE, 2009.

\bibitem{Fontugne_MAWILab_CoNEXT2010}
R.~Fontugne, P.~Borgnat, P.~Abry, and K.~Fukuda, ``{MAWILab}: Combining diverse
  anomaly detectors for automated anomaly labeling and performance
  benchmarking,'' in \emph{Proceedings of 6th International Conference on
  emerging Networking EXperiments and Technologies (CoNEXT 2010)}.\hskip 1em
  plus 0.5em minus 0.4em\relax ACM, 2010, p. Article No.\ 8.

\bibitem{NSL-KDDDataSet2009}
{Information Security Center of eXcellence (ISCX), UNB}, ``{UNB ISCX NSL-KDD}
  dataset,''
  \url{http://www.unb.ca/research/iscx/dataset/iscx-NSL-KDD-dataset.html},
  2009.

\bibitem{GaffneyUlvila2001IDSevaluation}
J.~E. {Gaffney Jr.} and J.~W. Ulvila, ``Evaluation of intrusion detectors: A
  decision theory approach,'' in \emph{Proceedings of 2001 IEEE Symposium on
  Security and Privacy}.\hskip 1em plus 0.5em minus 0.4em\relax IEEE, 2001.

\bibitem{Ali2013ATT4IDS}
M.~Q. Ali, E.~Al-Shaer, H.~Khan, and S.~A. Khayam, ``Automated anomaly detector
  adaptation using adaptive threshold tuning,'' \emph{ACM Transactions on
  Information and System Security}, vol.~15, no.~4, p. Article No.\ 17, 2013.

\end{thebibliography}
\fi


%\vfill


\end{document}